\documentclass{aa}
\usepackage[varg]{txfonts}
\usepackage{graphicx} 
\usepackage{xcolor}
\usepackage{hyperref}
\hypersetup{
    colorlinks=true,
    linkcolor=black,     
    urlcolor=blue,
    citecolor=blue
    }

\begin{document}
\title{A general relativistic magnetohydrodynamics extension to mesh-less schemes in the code GIZMO}
\titlerunning{GIZMO-GRMHD}
\author{Giacomo Fedrigo$^{1,2}$ and Alessandro Lupi$^{1,2,3}$}
\date{}

\institute{$^1$Como Lake centre for AstroPhysics, DiSAT, Università dell’Insubria, via Valleggio 11, 22100 Como, Italy \\$^2$INFN, Sezione di Milano-Bicocca, Piazza della Scienza 3, I-20126 Milano, Italy \\$^3$INAF - Osservatorio di Astrofisica e Scienza dello Spazio di Bologna, Via Gobetti 93/3, I-40129 Bologna, Italy}

\abstract
    {The profound comprehension of the evolution and phenomenology of an Active Galactic Nucleus requires an accurate exploration of the dynamics of the magnetized gaseous disk surrounding the massive black hole in the centre. Many numerical simulations have studied this environment using elaborate grid-based codes, but in recent years, new mesh-less schemes have exhibited excellent conservation properties and good accuracy at a more moderate computational cost. Still, none implement general relativistic magnetic fields, a fundamental ingredient to model an accretion disk around a massive black hole.
    We present here a general relativistic magnetohydrodynamics (GRMHD) scheme working within the mesh-less framework of the code \texttt{GIZMO}. We implement the hyperbolic divergence cleaning procedure, consistently extended to general relativistic effects, to keep the magnetic field divergence under safe levels. We benchmark the scheme against various relativistic magnetohydrodynamics stress tests, considering different dimensionalities and both a Minkowski or a Schwarzchild/Kerr background. To date, this is the first GRMHD scheme working in a mesh-free environment.}
 
\maketitle

\section{Introduction}\label{sec:Introduction}
In the last decades, the scientific community has witnessed a remarkable improvement in numerical techniques, which has been particularly useful in the astrophysical hydrodynamics scenario for the possibility of modelling the Universe with unprecedented accuracy and of numerically exploring previously inaccessible systems. Between the huge variety of techniques exploited in the astrophysical context, the most widely used surely are those relying on smoothed particle hydrodynamics \citep[SPH;][]{Lucy1977,GingoldMonaghan1977, Price2008, Price2012,CullenDehnen2010,ReadHayfield2012} or on grid based finite-volume methods \citep[e.g.][]{BergerColella1989,StoneMichael1992}. Both these approaches have their advantages and disadvantages; for example, while SPH codes have continuous adaptive resolution due to their Lagrangian nature, great conservation properties and provide a framework easily coupled with N-body gravity, they suffer from extra computational cost in neighbour finding and some resolution-scale noise associated with particle rearrangement that could cause fluid instabilities \citep{Lecoanet2016}.
 
On the other hand, grid based codes are numerically stable and provide good shock capture; however, they suffer from bad mass and angular momentum conservation, grid alignment effects, large artificial diffusion related to advection errors \citep{GardinerStone2005} and wasted computational time in empty regions, usually addressed with `adaptive mesh refinement' (AMR) implementations \citep{BergerColella1989}. Lately, a number of hybrid methods have tried to get the best of both approaches, as moving mesh schemes based on a Voronoi tessellation of the domain \citep{Springel2010,Pakmor2011}, Particle-in-Cell (PIC) simulations \citep{Aunai2024} and mesh-less methods \citep{LansonVila2008,LansonVila2008b}. The latter have been then implemented in astrophysical codes by \cite[][Weighted Particle MHD]{GaburovNitadori2011}, \citep[][\texttt{GIZMO}]{GizmoHD,GizmoMHD,Hopkins2016b}, and \citep[\texttt{OpenGADGET}]{opengadget24}. In particular, \texttt{GIZMO} has been extensively used in many astrophysical scenarios, ranging from the study of AGN accretion disks and circumbinary disks formation and evolution \citep{Franchini2022,Franchini2024,Mudit2024} to tidal disruption events \citep{Mainetti2017}, to star forming regions \citep{grudic2021,Lupi2021}, to cosmological galaxy formation simulations \citep{Lupi2017,Hopkins2018,Hopkins2022}, proving its accuracy and effectiveness.

All these inquiries were treated with Newtonian physics, yet only recently a general relativistic hydrodynamics (GRHD) extension has been introduced in \texttt{GIZMO} \citep{Lupi2023}. However, an implementation of general relativistic magnetohydrodynamics (GRMHD) in \texttt{GIZMO} was missing. Magnetic fields are ubiquitous in astrophysical plasma and many works studied the importance of highly magnetized regions in the proximity of compact objects. In facts, the magnetic field is a critical ingredient for the development of magnetorotational instabilities, potentially explaining the angular momentum transfer in AGN accretion disks \citep{BalbusHawley1991,Igumenshchev2003,Narayan2003}. In addition, magnetically dominated funnels close to the event horizon of single or binary black holes appear to be sources of relativistic jets of material, representing important features to look for in electromagnetic (EM) surveys \citep{Palenzuela2010,Giacomazzo2012,Bogdanovic2022,CattoriniGiacomazzo2024,Fedrigo2024}. Tidal disruption events are other important scenarios in which magnetic fields play an important role; in fact, when a star equilibrium gets destroyed by a close-by compact object, the initially confined magnetic field eventually permeates the inflowing stream of gas, affecting its dynamics and contributing to potential signature EM emissions. Finally, a series of works presenting a recent \texttt{GIZMO} simulation \citep{Hopkins2024,Hopkins2024a,Hopkins2024b,Hopkins2024c} studies the formation and evolution of magnetically dominated AGN accretion disks originating from cosmological initial conditions and resolving a dynamical range spanning from $\sim 100$ Mpc down to $\sim 100$ au in a consistent run. The accretion disk innermost region remains however unresolved and has been studied using the code \texttt{H-AMR} in \citet{Kaaz2024}.

These considerations motivate us to implement a GRMHD scheme in the \texttt{GIZMO} code. Although many other codes that evolve the GRMHD set of equations exist \citep{Gammie2003,DelZanna2007,GiacomazzoRezzolla2007,GRHydro,Etienne2015,Fragile2018,Kidder2017,Cipolletta2020,Liska2022}, none of them do it in a Lagrangian framework. To date, our scheme is the first one to solve the GRMHD equations exploiting the benefits of a mesh-less scheme. With our implementation, \texttt{GIZMO} can now self-consistently simulate the Universe from cosmological down to the black hole (BH) event horizon scales, including a plethora of physical prescriptions, such as: cosmological integration, dark matter, radiative heating and cooling \citep{Hopkins2017}, stellar and BH formation and feedback \citep[e.g.][]{grudic2021,Guszejnov2021,Hopkins2022,Bollati2024} chemistry, dust and now also GRMHD effects.

However, a solid implementation of magnetic fields represents a serious challenge due to the need to keep the magnetic field numerical divergence as low as possible. In fact, a non-zero $\nabla\cdot \textbf{B}$ can lead to wrong results or to code breaking; a relativistic scheme is even more sensible to this issue due to the necessity of retrieving the primitive physical quantities from conserved ones via a numerical iterative method (the conservative-to-primitive algorithm). It is easy to understand that a non-zero magnetic field divergence would break the convergence of the latter.

There are two methods commonly employed to control the spurious $\nabla\cdot \textbf{B}$ buildup: the constrained transport (CT) scheme \citep{EvansHawlet1988}, which is able to maintain $\nabla\cdot \textbf{B}$ constant to machine precision, and the evolution of the vector potential \citep{Etienne2012}. However, these methods are not straightforward to implement in a mesh-less scheme \citep{Tu2022}. Therefore, we opt here for a hyperbolic divergence cleaning approach \citep{Powell1999,Dedner2002}, taking inspiration from the Newtonian implementation already present in the original version of \texttt{GIZMO} \citep{GizmoMHD}. We show that our scheme can pass several special and general relativistic MHD stress tests, keeping the magnetic field divergence value under safe levels. 

This paper is organized as follows: in Sec.~\ref{sec:GRMHDEquations} we recall the GRMHD set of equations, written according to the 3+1 Valencia formulation; in Sec.~\ref{sec:NumericalImplementation} we present our numerical implementation of the scheme, quickly reviewing the \texttt{GIZMO} mesh-less method, and then focusing on the MHD additions to the GRHD implementation \citep{Lupi2023}; in Sec.~\ref{sec:DivergenceCleaning} we explain in detail our divergence cleaning approach; then, in Sec.~\ref{sec:CodeValidationTests} we show the results from several validation tests, both in special and general relativity. Finally, we draw our conclusions in Sec.~\ref{Sec:Conclusion}.

\section{GRMHD equations}\label{sec:GRMHDEquations}
In this section, we describe the GRMHD set of equations as implemented in our code. Our scheme follows the model described in the \texttt{GRHydro} \citep{GRHydro} code, implementing the Valencia formulation \citep{Banyuls1997} of the equations. We also refer the reader to \citet{Font2008} and \citet{MizunoRezzolla2024} for a review on numerical MHD and GRMHD and to \citet{Komissarov1999}, \citet{Koide1999}, \citet{Koide2000}, \citet{Koide2002}, \citet{DeVilliersHawley2003a}, and \citet{Gammie2003} for pioneering special and general relativistic MHD implementations in multidimensional simulation codes.
In what follows we use Greek indices for four-dimensional quantities, while Latin indices are used for three-dimensional ones. We also employ natural units, where $G=c=M_\odot=1$ and we make use of Einstein notation for repeated indices.

We assume a 3+1 spacetime decomposition in the Arnowitt-Deser-Misner (ADM) formalism \citep{ADM2008} as
\begin{equation}\label{eq:3+1Metric}
    ds^2 = g_{\mu\nu}dx^\mu dx^\nu = (-\alpha^2 +\beta_i \beta^i)dt^2 + 2\beta_i dx^i dt + \gamma_{ij}dx^i dx^j,
\end{equation}
where $g_{\mu\nu}$ is the 4-dimensional metric, $\alpha$ is the lapse function, $\beta^i$ is the shift vector and $\gamma_{ij}$ is the 3-dimensional spatial metric.
We concentrate on \textit{ideal GRMHD}, i.e. we assume infinite conductivity of the fluid and neglect viscosity and heat conduction. This corresponds to imposing $u_\nu F^{\mu\nu} = 0$, where $u_\nu$ is the four-velocity of the fluid and $F^{\mu\nu}$ is the Faraday tensor. Under this assumption, the electric field as seen by a Eulerian observer can be written as a function of the magnetic field $B^i$ as $E^i = -\alpha \varepsilon^{0ijk}v_j B_k$, while it vanishes in the fluid rest frame, as expected for a perfect conductor \citep{Banyuls1997,Anton2006,Font2008}.

To describe the fluid, we use the following quantities: $\rho$ is the fluid rest mass density, $\epsilon$ the specific internal energy, $P$ the gas pressure and $v^i = u^i/W + \beta^i/\alpha$ is the fluid three-velocity. In the latter we identify $W=(1-v^i v_i)^{-1/2}$ as the Lorentz factor, where $v_i v^i = \gamma_{ij} v^j v^i$. With the addition of the following definitions for the time and spatial components of the magnetic field in the fluid's rest frame, 
\begin{equation}\label{eq:Smallb}
\begin{split}
    &b^0 \equiv \frac{WB^j v_j}{\alpha},\\
    &b^i \equiv \frac{B^i}{W} + W(B^jv_j)\left (v^i - \frac{\beta^i}{\alpha}\right),\\
    &b^2 \equiv \frac{B^jB_j}{W^2} + (B^j v_j)^2,
\end{split}
\end{equation}
we can write the stress-energy tensor of ideal GRMHD as
\begin{equation}\label{eq:StressEnergyTensor}
\begin{split}
    T^{\mu\nu} &= \rho\left(1+\epsilon + \frac{(P+b^2)}{\rho}\right)u^\mu u^\nu + \left(P+\frac{b^2}{2}\right)g^{\mu\nu} - b^\mu b^\nu\\
    &\equiv \rho h^* u^\mu u^\nu + P^* g^{\mu\nu} -b^\mu b^\nu,
\end{split}
\end{equation}
where in the second line we have defined the magnetically modified enthalpy $h^* \equiv 1+\epsilon+(P+b^2)/\rho$ and the total pressure $P^* \equiv P + b^2/2$.

The equations of GRMHD are then derived from the mass density conservation law
\begin{equation}\label{eq:MassDensityConservation}
    \frac{1}{\sqrt{-g}}\partial_\mu(\sqrt{-g}\rho u^\mu)=0,
\end{equation}
where $g=$det$(g_{\mu\nu})$ is the determinant of the four-metric, the conservation of energy-momentum
\begin{equation}\label{eq:EnergyMomentumConservation}
    \nabla_\mu T^{\mu\nu}=0,
\end{equation}
and the Maxwell equations in the ideal regime
\begin{equation}\label{eq:MaxwellEquations}
    \nabla_\nu {}^*F^{\mu\nu}=0,
\end{equation}
where ${}^*F^{\mu\nu} = \frac{1}{2}\varepsilon^{\mu\nu\lambda\sigma}F_{\lambda\sigma}$ is the dual of the Faraday tensor.
This set of equations can be cast in hyperbolic form
\begin{equation}\label{eq:GRMHDHyperbolicForm}
    \frac{\partial \textbf{U}}{\partial t} + \frac{\partial \textbf{F}^i}{\partial x^i} = \textbf{S};
\end{equation}
here $\textbf{U}$ is the conserved quantities vector defined as
\begin{equation}\label{eq:ConservedVariables}
\textbf{U}=
\begin{cases}
    D &\equiv \sqrt{\gamma}\rho W, \\
    S_j &\equiv \sqrt{\gamma}(\rho h^*W^2 v_j - \alpha b^0b_j),\\
    \tau &\equiv \sqrt{\gamma}(\rho h^* W^2 - P^* - (\alpha b^0)^2) - D, \\
    \mathcal{B}^j &\equiv \sqrt{\gamma}B^j,
\end{cases}
\end{equation}
where $\gamma =$det$(\gamma_{ij})$ is the determinant of the spatial metric, \textbf{F} is the flux vector
\begin{equation}\label{eq:Fluxes}
\textbf{F}^i=
\begin{cases}
    &D\tilde{v}^i,\\
    &S_j\tilde{v}^i + \alpha\sqrt{\gamma}P^*\delta^i_j - \alpha b_j\mathcal{B}^i/W,\\
    &\tau\tilde{v}^i + \alpha\sqrt{\gamma}P^*v^i - \alpha^2 b^0\mathcal{B}^i/W,\\
    &\mathcal{B}^j\tilde{v}^i - \mathcal{B}^i\tilde{v}^j,
\end{cases}
\end{equation}
where $\tilde{v}^i = \alpha v^i - \beta^i$, and \textbf{S} is the sources vector \citep[]{Duez2005, ChangEtienne2020}
\begin{equation}\label{eq:Sources}
\textbf{S}=\alpha\sqrt{\gamma}
\begin{cases}
    &0,\\
    &\frac{1}{2}T^{\alpha\beta}\nabla g_{\alpha\beta},\\
    &(T^{00}\beta^i\beta^j + 2T^{0i}\beta^j+T^{ij})K_{ij} - (T^{00}\beta^i + T^{0i})\partial_i\alpha,\\
    &0,\\
\end{cases}
\end{equation}
where $K_{ij}$ is the extrinsic curvature of the metric. We note that the source terms depend on the derivative of the metric and we do not have any modification in form from the GRHD case (note however that the stress-energy also considers the magnetic energy-momentum contribution now).
\\Finally, the time component of Eq.s (\ref{eq:MaxwellEquations}) yields the divergence-free condition of the magnetic field
\begin{equation}\label{eq:DivergenceFreeB}
    \frac{1}{\sqrt{\gamma}} \partial_i \mathcal{B}^i =0;
\end{equation}
numerically, this constraint needs to be ensured during the system evolution. We enforce this condition with a `divergence cleaning' technique, which we describe in Sec. \ref{sec:DivergenceCleaning}.

An equation of state (EoS) is required to close the system of Eq.s (\ref{eq:GRMHDHyperbolicForm}). In our work, we always use a $\Gamma$-law EoS of the form $P=(\Gamma - 1)\rho \epsilon$, albeit our scheme also allows for different equations of state.

\section{Numerical implementation}\label{sec:NumericalImplementation}
In this section, we describe the numerical methods used to solve the GRMHD set of Eq.s (\ref{eq:GRMHDHyperbolicForm}), mainly focusing on the differences between the GRHD version of the scheme, presented in \citet{Lupi2023}, and the current implementation. After a brief recall of the \texttt{GIZMO} mesh-free methods, we describe the new volume evolution prescription, the GRMHD Riemann solver, and, finally, the methods used to ensure the divergence-free condition.  
\subsection{Review of the \texttt{GIZMO} mesh-less methods}\label{sec:GizmoReview}
The code \texttt{GIZMO} solves the evolution equations in a standard finite-volume Godunov-type style \citep{GizmoHD,GizmoMHD}. The unique trait of the code resides in the domain discretization: the volume is partitioned among elements (which we will call either `particles' or `cells' at need) according to a continuous weighting function $\Psi_i$, so that at each mesh-generating point position $\textbf{x}$, an infinitesimal volume $d\textbf{x}$ can be described by
\begin{equation}
    d\textbf{x} = \sum_i \Psi_i(\textbf{x},h(\textbf{x})),
\end{equation}
where $h(\textbf{x})$ is the kernel size. The latter is evaluated from the code for each particle to encompass a user-defined number of neighbors.
The weighting function $\Psi_i$ is defined as 
\begin{equation}\label{eq:WeightingFunction}
    \Psi_i(\textbf{x}, h(\textbf{x})) = \frac{W(\textbf{x}_i - \textbf{x} , h(\textbf{x}))}{\sum_j W(\textbf{x}_j - \textbf{x} , h(\textbf{x}))},
\end{equation}
where $W$ is any continuous, symmetric kernel function with compact support (for instance, we usually employ a cubic spline kernel function, unless otherwise indicated). The sum at the denominator normalizes the function $\Psi_i$ so that the volumes always sum correctly.

As in the GRHD case \citep{Lupi2023}, we decide to solve Eq.s (\ref{eq:GRMHDHyperbolicForm}) in the lab reference frame, as
\begin{equation}\label{eq:GRMHDEquationLabFrame}
    \frac{{\rm d} \textbf{U}}{{\rm d} t} + \nabla\cdot (\textbf{F} - \textbf{w} \otimes \textbf{U}) = \textbf{S},
\end{equation}
where $\mathbf{w}$ is the frame velocity.
As described in the original methods paper \citep{GizmoHD}, we can discretize Eq.s \eqref{eq:GRMHDEquationLabFrame} over the discrete elements defined above, by integrating over the domain volume, obtaining
\begin{equation}\label{eq:VolumeIntegrated}
    \frac{{\rm d}(V_i\textbf{U}_i)}{{\rm d}t} + \sum_j \textbf{\~{F}}_{ij} \cdot \textbf{A}_{ij} = V_i\textbf{S}_i,
\end{equation}
where $\textbf{\~{F}}_{ij}$ is the solution of the Riemann problem between the $i$-th particle and the $j$-th interacting neighbor, $V_i = \int \Psi_i(\textbf{x})d\textbf{x}$ is the particle volume and $\textbf{A}_{ij}$ is the normal vector to the `effective face area' between $i$-th and $j$-th interacting cells.

In principle, the frame velocity $\textbf{w}$ in Eq.s (\ref{eq:GRMHDEquationLabFrame}) can be chosen arbitrarily. \texttt{GIZMO} defines two different methods which depend on the assumption made for the frame velocities, as described in the original paper. One assumes the interacting face to be moving with the average fluid velocity of the interacting cells $w_{ij} = (\tilde{v}_i + \tilde{v}_j)/2$, defining the mesh-less Finite Volume (MFV) method, which resembles a standard moving-mesh finite volume scheme. The other assumes instead the frame velocity to be that of the contact discontinuity, i.e. the velocity which preserves the mass on both sides of the moving interface. This second choice defines the mesh-less Finite Mass (MFM) method, in which the particle masses are always preserved throughout the evolution of the system (more on that in Sec.~\ref{sec:RiemannSolver}). 

\subsection{MHD modifications}\label{sec:MHDModifications}
In this section, we describe the main method modifications from the GRHD version of the scheme, extensively described in \citet{Lupi2023}. The first difference resides in the implementation of the Riemann solver, which we describe in the following subsection.

\subsubsection{Riemann solver}\label{sec:RiemannSolver}
As in the GRHD case, we implement the one-dimensional HLL Riemann solver \citep{HLL}. 
We find that our single intermediate state HLL solver provides a computationally cheap and sufficiently accurate, although more diffusive, solution to the problem; in the future, we will also include a multi-state Riemann solver to the scheme, as the one by \citet{MignoneBodo2006} or \citet{MiyoshiKusano2005}, in order to correctly resolve contact discontinuities.

The fluxes given by the solver can be written as
\begin{equation}\label{eq:HLL}
    \textbf{\~F} = 
    \begin{cases}
        \textbf{F}_L - w^{\hat{n}}\mathcal{U}_L;  \ \ \ \ \ \ \ \ \ \ \ \ \ \ \ \ \ \ \ \  w^{\hat{n}} < \lambda_{min}\\
        \textbf{F}_{HLL} - w^{\hat{n}}\mathcal{U}_{HLL}; \ \ \ \ \ \  \lambda_{min} \leq  w^{\hat{n}} \leq \lambda_{max}\\
        \textbf{F}_{R} - w^{\hat{n}}\mathcal{U}_{R}; \ \ \ \ \ \ \ \ \ \ \ \ \ \ \ \ \ \ \ \  w^{\hat{n}} > \lambda_{max}\\
    \end{cases}
\end{equation}
where $\mathcal{U}_{HLL}$ are the intermediate states evaluated as
\begin{equation}\label{eq:HLLState}
    \mathcal{U}_{HLL} = \frac{\lambda_{max} \mathcal{U}_R - \lambda_{min}\mathcal{U}_L + \textbf{F}_L - \textbf{F}_R}{\lambda_{max}-\lambda_{min}}
\end{equation}
and $\textbf{F}_{HLL}$ are the corresponding fluxes
\begin{equation}\label{eq:HLLFluxes}
    \textbf{F}_{HLL} = \frac{\lambda_{max}\textbf{F}_L - \lambda_{min}\textbf{F}_R +\lambda_{max}\lambda_{min}(\mathcal{U}_R-\mathcal{U}_L)}{\lambda_{max}-\lambda_{min}}.
\end{equation}
In these equations, $w^{\hat{n}}$ is the face velocity perpendicular to the interface, whereas $\lambda_{min}$ and $\lambda_{max}$ are the slowest and the fastest wave speeds of the Riemann problem, i.e. the magnetosonic speeds. We evaluate such velocities by solving the quadratic dispersion relation
\begin{equation}\label{eq:DispersionRelation}
\begin{split}
    \lambda^2[W^2(V^2-1) - V^2] -2\lambda[W^2 \tilde{v}^{\hat{n}}(V^2 - 1) + V^2\beta^{\hat{n}}] + \\ [(W \tilde{v}^{\hat{n}})^2(V^2-1) + V^2(\alpha^2\gamma^{\hat{n}\hat{n}} - \beta^{\hat{n}}\beta^{\hat{n}})]=0
\end{split}
\end{equation}
where $V^2 \equiv v_A^2 + c_s^2(1-v_A^2)$, with $v_A = \sqrt{b^2/(\rho h^*)}$ the Alfvén velocity and $c_s$ the fluid sound speed.

In order to guarantee the stability of the scheme, the Courant-Friedrisch-Levy (CFL) condition must be fulfilled during the calculation. Even though the solution of $\lambda$ should be used in this case, for computational efficiency we employ $V$ as the maximum signal velocity. This represents an approximate evaluation of the signal speed, but gives a good enough estimate for our purpose.

To conclude this subsection, we recall that when \texttt{GIZMO} is running in MFM mode, the face velocity $w^{\hat{n}}$ is assumed to be moving with the zero-mass flux velocity, which for our HLL solver is defined as
\begin{equation}\label{eq:MFMFrameVel}
    w^{\hat{n}} = \frac{\textbf{F}_{D,HLL}}{\mathcal{D}_{HLL}}.
\end{equation}

\subsubsection{Time integration and volumes evolution}\label{sec:VolumeEvolution}
After the flux computation, we need to evolve the conserved variables in time according to the generalized leap-frog scheme, as described in \citet{Lupi2023}. The only difference with the GRHD scheme lies in the explicit update of the conserved variables through Eq.s (\ref{eq:VolumeIntegrated}), both for the "kick" procedure on active particles over half a timestep $\Delta t /2$ and the prediction operation over a full timestep $\Delta t$. In fact, in the GRHD scheme, all the conserved quantities $(D, S_j, \tau)_i$ are proportional to the particle density $D_i$, i.e. proportional to the particle mass $m_i$ once volume-integrated. This results in just a normalization factor in the conservative-to-primitive inversion algorithm, irrelevant to its solution (Sec. \ref{sec:C2PConversionWvReconstruction}). However, in the volume-integrated GRMHD conserved variables, a proportionality to $V_i^2$ appears in the terms containing $\mathcal{B}_i^2$ in the momentum and energy density definitions, breaking the inversion scheme. For this reason, we modify our scheme introducing an explicit evolution of the volume $V_i$, as 
\begin{equation}\label{eq:TimeIntegration}
    \textbf{U}_i^{t+\Delta t} = \frac{V_i^t\left[\textbf{U}_i^t + \textbf{S}(\textbf{U}^t_i) \Delta t \right]- \sum_j\textbf{A}_{ij}^t\textbf{\~F}^t_{ij}\Delta t }{V_i^{t+\Delta t}}.
\end{equation}
This guarantees that the local conservative quantities are consistently passed to the conservative-to-primitive solver.

This procedure boils down to how to explicitly compute the updated volume $V_i^{t+\Delta t}$. We decided to estimate it by exploiting the continuity equation as $V_i^{t+\Delta t} = V^t_i \exp(\dot{V}_i\Delta t/V^t_i)$, where $\dot{V}_i = \sum_j \textbf{A}_{ij}\cdot (\mathbf{v}_{part,i} - \mathbf{w}_{ij})$; here $\textbf{v}_{part,i}$ is the i-th particle velocity (which is equal to the fluid velocity in the MFM scheme), $\textbf{w}_{ij}$ the velocity of the Riemann problem interface\footnote{We note that, in order to be consistent throughout the whole evolution, $\textbf{w}_{ij}$ must be the frame velocity actually used in the Riemann problem, i.e. it must be evaluated according to Eq. (\ref{eq:MFMFrameVel}) in MFM mode.}, and $\textbf{A}_{ij}$ is the effective face area between particles $i$ and $j$.\footnote{Actually, in the second half-step kick we start from $V_i^{t+\Delta t} = M_i^{t+\Delta t} / D_i^{t + \Delta t}$ at the end of the step and we estimate $V_i^{t+\Delta t/2} = V_i^{t+\Delta t} \exp (-\dot{V}_i\Delta t/V^{t+\Delta t}_i)$ back at the beginning of the half-kick.} For consistency, the same prescription is employed in the density prediction step, i.e. $D^{t+\Delta t } = D^t \exp(-\dot{V}\Delta t/V^t)$.

\subsubsection{Other code modifications}\label{sec:C2PConversionWvReconstruction}
In this section, we discuss other code modifications implemented in our GRMHD extension of \texttt{GIZMO}.
\begin{enumerate}
    \item Despite the use of slope limiters, the process of reconstructing the primitive velocity at the position of the face might yield a resulting fluid velocity slightly larger than unity. In the GRHD version of the code, this was prevented by means of a user-defined limiter on the maximum Lorentz factor allowed by the code. Here instead, we have found more convenient to reconstruct the quantity $Wv^i$ at the face position, as this quantity has physical values ranging in $(-\infty, \infty)$, thus naturally avoiding causality violation. After reconstruction, the corresponding Eulerian $v^i$ is then easily recovered \citep{Noble2006} and passed to the Riemann solver. Note that the implementation of a limiter in the evaluation of the Lorentz factor is still available in the code, but it is now optional, and not employed in any of the tests presented in this work.
    \item Because of the presence of the Lorentz factor in the definition of the conservatives variables, primitive quantities cannot be recovered analytically, but only through iterative numerical techniques for the inversion of Eq.s (\ref{eq:ConservedVariables}). Here, we implement the 2D inversion scheme by \citet{Noble2006}, suitably adapted to our scheme, which recovers the GRMHD primitive variables via a Newton-Raphson method. This scheme considers the presence of magnetic fields, can be easily extended to generic EoS and provides a stable and computationally efficient solution to this inversion problem. We note that keeping the divergence of the magnetic field low is crucial for the correct behaviour of the conservative-to-primitive solver.
    \item In some pathological cases, when the magnetic field contributes to a significant fraction of the total energy of the fluid, the recovery of the specific internal energy $\epsilon$ from the conserved energy $\tau$ can become a non-trivial numerical task, yielding significant errors on the evolution of the primitive quantities. In our tests, we encountered this problem only in the magnetized Bondi accretion test, when dealing with a strong gravitational field (Sec. \ref{sec:BondiAccretion}). For this reason, we implement an optional energy-entropy switch, as suggested in \citet{DelZanna2007}. When active, we evaluate the entropy function $s\equiv P/\rho^\Gamma$, which we evolve through the conservation equation
\begin{equation}\label{eq:EntropyAdiabaticEquation0}
    \nabla_\mu(\rho su^\mu)=0,
\end{equation}
    that can be rewritten as
\begin{equation}\label{eq:EntropyAdiabaticEquation}
    \partial_t(Ds)+\partial_i(Ds\tilde{v}^i)=0.
\end{equation}
    We note that the last equation only holds for adiabatic flows, in the absence of shocks or any other energy dissipation mechanisms. During the fluid evolution, for each particle, the code checks whether the magnetic field contributes to a significant fraction of the total energy and if any shock is present in its interaction with neighbors particles.\footnote{If the local fluid velocity is greater than the local sound speed, the code assumes that the interaction produces a shock.} If the first condition is met and shocks are absent, the internal energy is directly recovered from the entropy $s$. Otherwise, if the first condition is false or a shock is present in any direction, we evaluate the internal energy relying on the standard evolution Eq.s \ref{eq:GRMHDHyperbolicForm}).
\end{enumerate}

\subsection{Divergence cleaning}\label{sec:DivergenceCleaning}
The ideal GRMHD Eq.s (\ref{eq:GRMHDHyperbolicForm}) are derived using Eq.~(\ref{eq:DivergenceFreeB}), i.e. the no-monopole constrain. It is well established, however, that such constraint is not always guaranteed in numerical schemes. Indeed, the numerical approximations performed during the calculations can easily result in the divergence of the magnetic field significantly differing from zero, thus leading to an incorrect and unphysical evolution of the conserved quantities or, worse, to code breaking. Keeping the emerging errors in Eq.~\eqref{eq:DivergenceFreeB} under control is therefore crucial for a correct evolution of the systems considered. This is usually achieved with numerical techniques such as the constrained transport \citep[CT;][]{EvansHawlet1988} or the vector potential evolution \citep{Etienne2012} methods. While successful implementations of these methods in moving-mesh relativistic MHD schemes exist \citep[e.g.][]{HeTang2012,Fragile2018}, we opt here for a divergence cleaning technique, to maintain consistency with the Newtonian MHD version of the code \citep{GizmoMHD}. To date, this is the first implementation of a divergence cleaning method in a GRMHD set of equations solved on a mesh-less scheme.

We now describe our divergence cleaning implementations, starting from the Powell's ``8-wave'' cleaning prescription. As described in the original paper \citep{Powell1999}, a set of source terms is needed to stabilize the MHD equations against the development of spurious magnetic divergence when the fluxes are evaluated. A special relativistic version of this scheme is analyzed in depth by \citet{WuShu2021}. Here, we write a general relativistic analogue, similar to the one presented in \citet{Liebling2010}.

We extend the set of Eq.s (\ref{eq:GRMHDHyperbolicForm}) with additional source terms in the form
\begin{equation}\label{eq:PowellSources}
    \textbf{S}_{Powell} = - \alpha (\partial_i \mathcal{B}^i)
    \begin{cases}
        0,\\
        \frac{B_j}{W^2} + v_jB^kv_k, \\
        B^kv_k,\\
        v^j-\frac{\beta^j}{\alpha}.
    \end{cases}
\end{equation}

The critical point here resides in the need to consistently evaluate the $\partial_i\mathcal{B}^i$ term. As also reported in the original MHD \texttt{GIZMO} paper (and suggested by \citealt{GaburovNitadori2011}), the only valuable choice is to compute the divergence term directly from the Riemann solver outputs as
\begin{equation}\label{eq:DivBEvaluation}
    (V \partial_k\mathcal{B}^k)_i = -\sum_j \bar{\mathcal{B}}^\perp_{ij}|\textbf{A}_{ij}|,
\end{equation}
where 
\begin{equation}\label{eq:BNormal}
    \bar{\mathcal{B}}^\perp_{ij} = \frac{\mathcal{B}^\perp_i + \mathcal{B}^\perp_j}{2}
\end{equation}
is the average magnetic field perpendicular to the face between each $i$-$j$ interacting pair of particles.
This prescription alone can complete all of our tests without the code breaking. However, some test problems (e.g. the magnetic rotor, Sec. \ref{sec:MagneticRotor}) yield wrong solutions. In order to address these inconsistencies and reduce the errors in the solutions, we add a hyperbolic divergence cleaning on top of the Powell's prescription. Presented for the first time in \citet{Dedner2002} in its Newtonian formulation and later extended to special and general relativity \citep[]{Neilsen2006,Anderson2006,Liebling2010,Penner2011,GRHydro,Fambri2018}, the method consists in the addition of a scalar field $\psi$ used to advect divergence errors away from their source location as fast as possible while damping them.

To consistently derive the equations governing the evolution of the $\psi$ field, we start from a modification of Maxwell's equation (following \citealt{GRHydro})
\begin{equation}\label{eq:DednerMaxwellEquations}
    \nabla_\mu\left({}^*F^{\mu\nu} + g^{\mu\nu}\hat{\psi}\right) = \sigma n^\nu c_h\hat{\psi},
\end{equation}
where $\hat{\psi}=\psi/c_h$, $n^\nu = (1/\alpha,-\beta^j/\alpha)$ is the normal vector to the hypersurface of the 3+1 formulation of spacetime and where we have introduced the advection velocity $c_h$ and the damping factor $\sigma$. As suggested in \citet{TriccoPrice2016}, we follow the $\hat{\psi}$ quantity to take in consideration the variability of the advection velocity $c_h$, obtaining a better evolution of the energy of the system.
We can now solve Eq.s (\ref{eq:DednerMaxwellEquations}) for $\nu=j$ (spatial components) and $\nu=0$ (time component), obtaining
\begin{equation}\label{eq:DednerCleaningBase}
\begin{cases}
    \partial_t \mathcal{B}^j + \partial_i(\tilde{v}^i\mathcal{B}^j -\tilde{v}^j\mathcal{B}^i) = -\alpha\sqrt{\gamma}\gamma^{ij}\partial_i\hat{\psi}+\beta^j\partial_i\mathcal{B}^i,\\
    \partial_t \hat{\psi} - \beta^i \partial_i\hat{\psi} = -\alpha \gamma^{
    -1/2}\partial_i\mathcal{B}^i - \alpha\sigma c_h \hat{\psi}.
\end{cases}
\end{equation}

As suggested in \citet{Dedner2002} and implemented in the MHD version of \texttt{GIZMO}, we consider an additional advection term $\alpha v^i\partial_i\hat{\psi}$ in the scalar field $\hat{\psi}$ evolution equation, which gives
\begin{equation}\label{eq:DednerCleaning}
\begin{cases}
    \partial_t \mathcal{B}^j + \partial_i(\tilde{v}^i\mathcal{B}^j -\tilde{v}^j\mathcal{B}^i) = -\sqrt{\gamma}(\beta^jv^i+\alpha\gamma^{ij})\partial_i\hat{\psi}+\beta^j\partial_i\mathcal{B}^i,\\
    \partial_t \hat{\psi} + \tilde{v}^i \partial_i\hat{\psi} = -\alpha \gamma^{-1/2} \partial_i \mathcal{B}^i - \alpha\sigma c_h\hat{\psi}.
\end{cases}
\end{equation}
If we now multiply the second equation  by the conserved density $D$ and we rearrange the terms using the mass continuity Eq. (\ref{eq:GRMHDHyperbolicForm}), we can rewrite the evolution equation for $\hat{\psi}$ as
\begin{equation}\label{eq:DednerDBased}
    \partial_t (D\hat{\psi}) + \partial_i(\tilde{v}^iD\hat{\psi}) = -\frac{\alpha D \partial_i \mathcal{B}^i}{\sqrt{\gamma}} - \alpha\sigma c_hD\hat{\psi}.
\end{equation}
Once averaged over the volume $V_i$, Eq. (\ref{eq:DednerDBased}) yields the evolution of the mass-based $\hat{\psi}$ field as implemented in our code, which translates into a much easier evaluation of the flux term.

\begin{figure*}
\begin{center} 
\includegraphics[width=.9\textwidth]{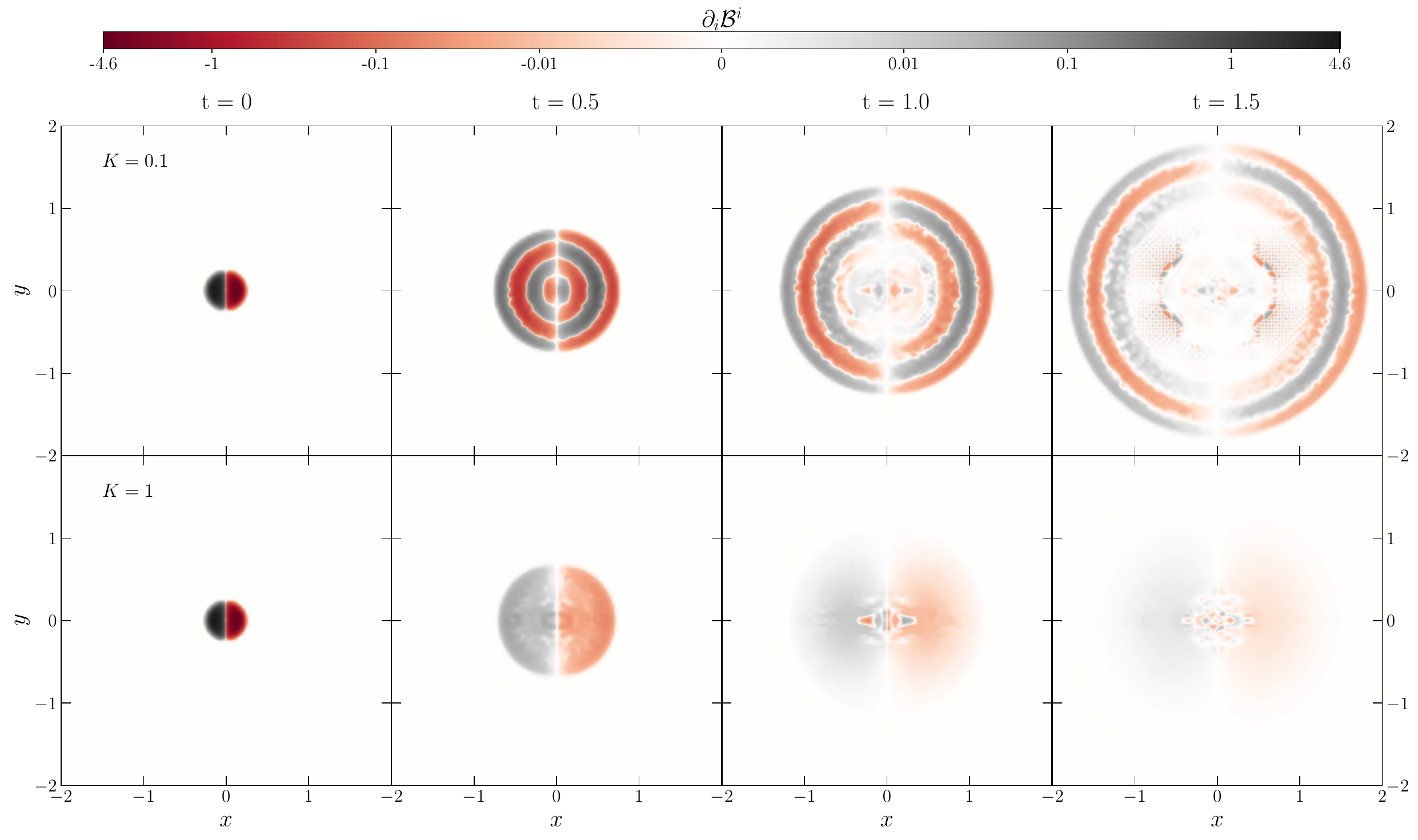}
\end{center}
\caption{Evolution of the monopole test. Two-dimensional slice at $z=0$ of the magnetic field divergence at the initial time (left) and at three later times. Top panels display the solution with damping parameter $K=0.1$ and the bottom panels with $K=1$.}\label{fig:Monopole}
\end{figure*}

As done for the $\partial_i\mathcal{B}^i$ term in Eq. \eqref{eq:BNormal}, we need to evaluate the gradient of $\hat{\psi}$ in a consistent way. In our implementation, this is done by considering the Rusanov fluxes between each pair of interacting particles
\begin{equation}\label{eq:RusanovFluxes}
    \textbf{F}_{HLL} = \frac{1}{2}(\textbf{F}_L + \textbf{F}_R) - \frac{\tilde{\lambda}}{2}(\mathcal{U}_R-\mathcal{U}_L),
\end{equation}
for the decoupled ($\mathcal{B}^\perp$, $\hat{\psi}$) system, where $\tilde{\lambda}$ is the eigenvalue of the two-dimensional problem. Here, $\tilde{\lambda}=\alpha\sqrt{\gamma^{\perp\perp}}$, where $\gamma^{\perp\perp}$ is the spatial metric component perpendicular to the interface, and from Eq.~(\ref{eq:RusanovFluxes}) we obtain
\begin{equation}\label{eq:PsiNormalBNormal}
    \begin{cases}
        \bar{\mathcal{B}}^\perp_{ij} = \frac{1}{2}(\mathcal{B}^\perp_L + \mathcal{B}^\perp_R) - \frac{1}{2}\sqrt{\gamma\gamma^{\perp\perp}}(\hat{\psi}_R - \hat{\psi}_L),\\
        \bar{\psi}_{ij}=\frac{1}{2}(\hat{\psi}_L+\hat{\psi}_R) - \frac{1}{2}(\gamma\gamma^{\perp\perp})^{-1/2}(\mathcal{B}^\perp_R - \mathcal{B}^\perp_L).
    \end{cases}
\end{equation}
Finally, $\bar{\mathcal{B}}^\perp_{ij}$ and $\bar{\psi}_{ij}$ are used to evaluate the divergence of $\mathcal{B}^j$ with Eq. (\ref{eq:DivBEvaluation}) and the gradient of $\hat{\psi}$ as
\begin{equation}\label{eq:GradPsiEvaluation}
    (V\partial_k\hat{\psi})_i = - \sum_j \bar{\psi}_{ij}\textbf{A}_{ij}.
\end{equation}

In principle, we are free to choose the Dedner parameters $c_h$ and $\sigma$ arbitrarily. After extensive tests of the scheme, we have opted for defining $\sigma=K/\Delta x$, with $K=0.1$ (except for the blast wave tests where we use $K=0.75$), where $\Delta x$ is the inter-particle spacing/effective cell size. In a relativistic scheme, the advection velocity is usually set equal to the speed of light, i.e. $c_h=1$, since this is the largest value permitted that does not violate causality. This choice, however, would break the hierarchical time-stepping in \texttt{GIZMO}, thus requiring the entire system to advance at the smallest possible pace. In our implementation, instead, we decided to employ the local magnetosonic speed $v_{ms}$, similar to what is done in the Newtonian MHD version of the code, suitably augmented by a user-defined factor $f$, i.e. $c_h = v_{ms}(1 + f)/(1+fv_{ms}^2)$.\footnote{Note that the augmented speed is simply the relativistic composition of $v_{ms}$ with itself multiplied by $f$.} In our tests, we typically assume $f=1$ (except for the magnetized TOV equilibrium test where we use $f=2.5$), which represents a compromise between computational efficiency and an effective divergence cleaning.\footnote{Note that the user is allowed to force $c_h=1$ at compile-time. In our tests, this choice yields the best cleaning of $\partial_i\mathcal{B}^i$, but negligible differences in the evolution of the conserved quantities.} We note that, while $c_h$ in Eq.s (\ref{eq:PsiNormalBNormal}) is evaluated from the local quantities entering the Riemann problem for the considered pair of particles, the maximum among all $c_h$ values from the interacting particles pair is employed in Eq.s (\ref{eq:DednerCleaning})-(\ref{eq:DednerDBased}).
We also note that when hyperbolic divergence cleaning is employed, the additional advection speed must be considered when evaluating the system time-step through the CFL condition.

Finally, in order to stabilize the scheme, we need to include two additional source terms in the evolution of the momentum density $S_j$ and energy density $\tau$, yielding a Dedner source vector of the form
\begin{equation}\label{eq:DednerStabilityTerms}
S_{Dedner}=
\begin{cases}
    0,\\
    \Xi^k \left[2B_kv_j - B_jv_k -\gamma_{kj}(B^iv_i)\right],\\
    \Xi^k \left[2B_k (1-\frac{1}{2W}) - v_k(B^i v_i) \right],\\
    \Xi^j,
\end{cases}
\end{equation}
where
\begin{equation}\label{eq:Xi}
    \Xi^k \equiv \left[-\sqrt{\gamma}(\beta^kv^i+\alpha\gamma^{ki})\partial_i\hat{\psi}+\beta^k\partial_i\mathcal{B}^i\right].
\end{equation}

This concludes the description of our divergence cleaning procedure.

\section{Code validation tests}\label{sec:CodeValidationTests}
We now test the robustness and accuracy of our implementation against some standard tests, both in special and general relativistic MHD. Unless otherwise indicated, all the tests are run with the complete divergence cleaning scheme (with parameters $K=0.1$ and $f=1$).
\subsection{Special relativistic magnetohydrodynamics}\label{sec:SRTests}
\subsubsection{Monopole}\label{sec:Monopole}

To test the behaviour and the effectiveness of our divergence cleaning prescription, we perform a simple test where we initialize a magnetic monopole in a static fluid. The initial setup is similar to the one presented in \citet{GRHydro}: a distribution of $100^3$ particles initially placed on a Cartesian mesh fills a periodic cubic box of side length 4. The fluid is uniform in density and pressure and it is initially at rest. The magnetic field is set equal to zero everywhere except in a central region where a monopole is introduced in the form
\begin{equation}\label{eq:Monopole}
    B^x = 
    \begin{cases}
    e^{-r^2/R_c}-e^{-1};\ \ \ \ &r<R_c   \\
    0;  &r \geq R_c.
    \end{cases}
\end{equation}
where $R_c=0.2$. The adiabatic index $\Gamma$ is set to $5/3$ during the evolution, we assume a Minkowski flat background and we choose the Dedner parameter to be $c_h=1$ and either $K=(0.1, 1)$. We report the results of the test performed in MFM mode but we find no difference in the MFV case.

In Fig. \ref{fig:Monopole} we show a two-dimensional slice of the magnetic field divergence evaluated according to Eq.~ (\ref{eq:DivBEvaluation}) at four times, $t=0,0.5,1,1.5$, for both the $K=0.1$ and the $K=1$ cases. Our scheme can correctly remove spurious monopoles, advecting and damping them in a relatively short time. As expected, the $K=1$ run displays a stronger reduction of $\partial_i \mathcal{B}^i$; however, a slightly slower suppression of $\hat{\psi}$ is recommended so that the evolution of the magnetic field itself is better affected by the corrective terms in Eq. (\ref{eq:DednerCleaning}). This analysis and the following tests show that higher $K$ values result in inefficient cleaning while too low $K$ values lead to system instabilities. We also note that even in this simple test, the corrective terms in Eq.s (\ref{eq:DednerStabilityTerms}) are necessary to keep the other evolved quantities stable.
For completeness, in Appendix \ref{AppendixB} we show results of the same test performed with the Powell scheme only.

\subsubsection{MHD shocktubes}\label{sec:Shocks}

\begin{figure*}
\begin{center} 
\includegraphics[width=.75\textwidth]{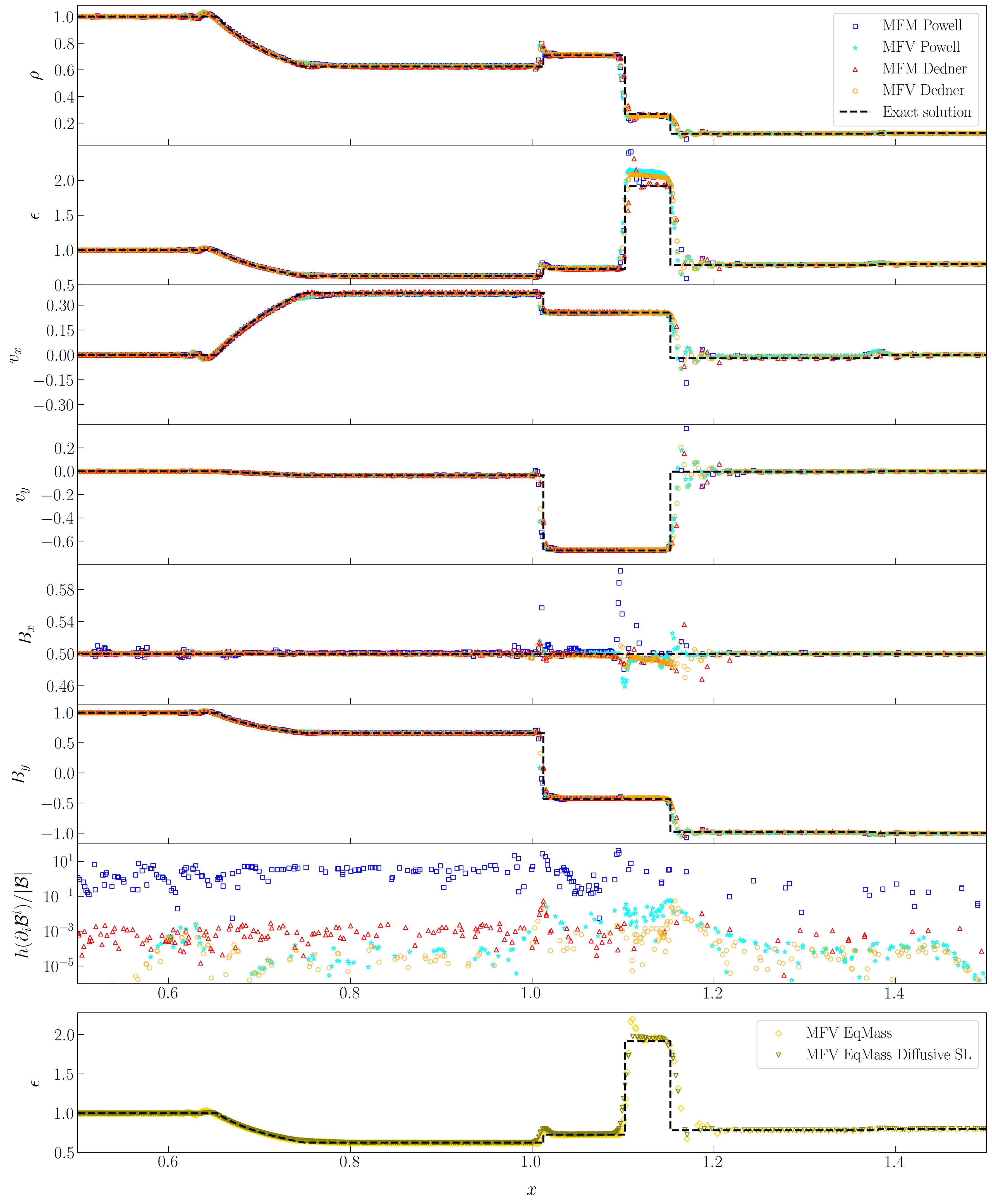}
\end{center}
\caption{Slice at $y=y_0$ of primitive quantities at time $t=0.4$ of the Balsara1 test. The magnetic field divergence in units of the magnetic field intensity is plotted in the seventh panel. Blue squares (Powell) and red triangles (Dedner) mark the MFM solutions, while cyan stars (Powell) and orange circles (Dedner) indicate the MFV ones. In the bottom panel, we display the specific internal energy when the problem is initialized with equal-mass particles and evolved with the MFV scheme; gold diamonds mark the solution employing standard slope limiters, while olive triangles show the results using more diffusive slope limiters. The correct solution, evaluated through the exact Riemann Solver by \citet{GiacomazzoRezzolla2006}, is plotted with a black dashed line.}\label{fig:Balsara1}
\end{figure*}

We report here the results of two one-dimensional special relativistic MHD tests. Presented for the first time in \citet{Balsara2000}, these tests consist of two-state (left and right) Riemann problems, particularly useful to demonstrate the ability of the code to capture shocks. In particular, we performed the relativistic analogue of the Brio-Wu shock tube problem \citep{BrioWu1988}, usually called ``Balsara1'' test, and a relativistic MHD collision, called ``Balsara4'' test.

The initial conditions of the Balsara1 problem consist of a left state $(\rho,v^x,v^y,v^z,\epsilon,B^x,B^y,B^z)_L=(1.0,0,0,0,1.0,0.5,1.0,0)$ and a right state $(\rho,v^x,v^y,v^z,\epsilon,B^x,B^y,B^z)_R=(0.125,0,0,0,0.8,0.5,-1.0,0)$, with $\Gamma = 2$. The discontinuity exists in the x-direction. The domain is a periodic 2D box of length $2$ along the $x$ direction and $0.05$ along the y-direction. 

To perform the test in the MFM mode, we initialize a distribution of equal-mass particles placed on a Cartesian grid, so that 2297 particles are sampling the x-direction. To determine the two different states' rest-mass densities, the left state has a particle number density eight times higher than the right. Because of periodicity, two discontinuities are present, one at the centre of the box, and one at the edge. We focus here only on the central one, as the other is simply a mirrored version of it, hence we limit our "active" region to the central half of the box. This translates in a number of "active" particles along the x-direction of $\sim1148$. For the MFV case, we discretize the fluid with a close packed lattice with 2000 particles along the x-direction (1000 "active" particles).

\begin{figure*}
\begin{center} 
\includegraphics[width=.75\textwidth]{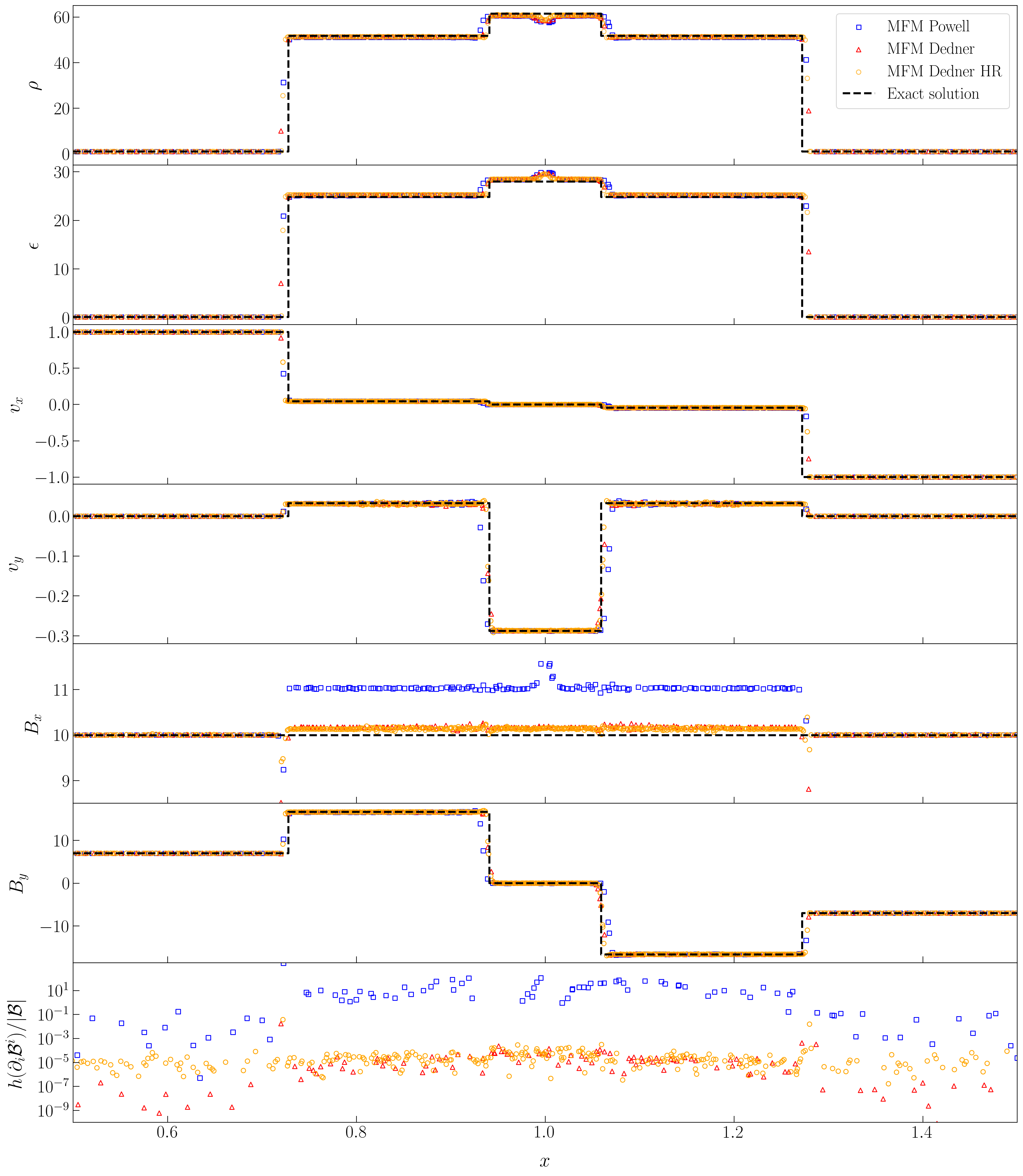}
\end{center}
\caption{Slice at time $t=0.4$ and constant $y$ of the primitive quantities from the Balsara4 test performed in the MFM mode. The magnetic field divergence in units of its intensity is plotted in the bottom panel. Blue squares mark the solution computed with Powell terms only, while red triangles indicate the one computed with the complete divergence cleaning scheme. The high resolution run is displayed with orange circles. Quantities $v_z$ and $B_z$ are not plotted but exhibit analogous behaviors to $v_y$ and $B_y$. The correct solution, evaluated through the exact Riemann Solver by \citet{GiacomazzoRezzolla2006}, is plotted with a black dashed line.}\label{fig:Balsara4}
\end{figure*}

In Fig.~\ref{fig:Balsara1}, we plot our solution of the Balsara1 test at $t=0.4$ performed in both MFM and MFV modes, using Powell terms only and with the complete cleaning scheme (Powell+Dedner). In the latter, we employ $K=0.1$ and $f=1$. 
The exact solution reported in the panels is computed with the exact RMHD Riemann Solver by \citet{GiacomazzoRezzolla2006}. Our scheme is able to reproduce the correct solution with good accuracy. We recognize some small oscillations on the left of the rarefaction fan and on the right of the fast magnetosonic wave front in all our runs; these artifacts can be controlled with more diffusive slope limiters when performing the reconstruction of the primitive quantities on the interacting faces. In this test, the specific internal energy $\epsilon$ is slightly overestimated at the contact discontinuity; close to the discontinuity, this effect is more severe in the MFM case, but exhibits an overall higher value in the MFV one. At the shock front (at $x\simeq1.15$) oscillations develop in the velocity and the specific internal energy evolution; this effect is probably related to the high volume compression the particles witness when the shock front is approaching. We can attenuate this artifact with more stringent slope limiters, lower order reconstructions of the primitive quantities, by employing bigger particle kernels (e.g. the Wendland C4 kernel function) or going to higher resolution, but some oscillations are always present when particle motion is allowed. Furthermore, initializing a particle configuration without sharp mass gradients relieves these difficulties.

In the bottom panel of Fig.~\ref{fig:Balsara1}, we display the specific internal energy when the problem is initialized with equal-mass particles and evolved with the MFV scheme. When compared with the same problem initialized with equal-volume particles (orange circles in the second panel of Fig.~\ref{fig:Balsara1}), the quantity $\epsilon$ exhibits a better agreement with the exact solution in the region between the contact wave and the right-going slow shock, resembling the MFM solution. Finally, in the same panel, we show results obtained using more stringent slope limiters; while being more diffusive, the solution has significantly fewer numerical artifacts; in fact, both the oscillations on the fast wave and the right slow wave, and the overestimation of $\epsilon$ in front of the contact discontinuity are attenuated.

In a planar shock test, the $\mathcal{B}^x$ evolution is a direct indication of the ability of the code to maintain a low magnetic field divergence, as it should remain constant in space and time. We recognize a good evolution of this quantity in all our runs, with only a moderate spike at the contact discontinuity in the MFM mode; however, with the addition of the Dedner divergence cleaning, the $\mathcal{B}^x$ evolution is improved overall. Finally, in the MFM runs, the magnetic field witnesses a slow noise build-up in regions yet unaffected by the shock; this is due to the volume discretization of the domain being performed on a 2D Cartesian grid in our initial conditions, which is not optimal for our $\partial_i \mathcal{B}^i$ and flux evaluations. Indeed, a better mapping of the simulated volume via a closed packed lattice, as done in the MFV case, completely removes this artifact.\footnote{Note that it is non-trivial to initialize an equal-mass particle distribution with two different density states using a closely packed lattice without introducing numerical artifacts at the discontinuity. Therefore, we opted for a simpler Cartesian distribution.} As a check of our implementation, we also performed this test with a fixed particle distribution (similar to a classical finite-volume scheme on a grid). The results, despite not shown, are in perfect agreement with the exact solution, without any of the numerical artifacts mentioned above, suggesting that their origin has to be ascribed to the particle motion, which reduces numerical diffusivity at relatively low resolution, but at the same time introduces some spurious oscillations \citep{GizmoHD}.

For the Balsara4 problem, we initialize a left-state $(\rho,v^x,v^y,v^z,\epsilon,B^x,B^y,B^z)_L=(1.0,0.999,0,0,0.15,10,7.0,7.0)$ and a right-state $(\rho,v^x,v^y,v^z,\epsilon,B^x,B^y,B^z)_R=(1.0,-0.999,0,0,0.15,10,-7.0,-7.0)$ with $\Gamma=5/3$. Since the initial Lorentz factor is $W=22.366$, this test is highly relativistic. We fill a two-dimensional $(3,0.05)$ periodic box with a closely packed lattice so that there are 500 ``active'' particles sampling the $x$-direction.

In Fig. \ref{fig:Balsara4}, we plot our solution at time $t=0.4$ performed in MFM mode, using either the Powell terms only or with the complete cleaning scheme, alongside the exact solution of the problem. Quantities $v_z$ and $B_z$ are not plotted but exhibit analogue behaviours to $v_y$ and $B_y$. Our code is able to correctly compute the evolution of the quantities, with only a small overestimation of the specific internal energy and an underestimation of the rest mass density (yielding a correct evaluation of the pressure) in the central regions. The positions of the shock fronts are slightly off, becoming more accurate with the Dedner cleaning scheme. The outer shocks remain instead slightly offset, because of the poor sampling around the strong density jump, which is reflected in a smoothing of the discontinuity across a few neighbors. To verify this hypothesis, we performed a high resolution run with $\sim1166$ active particles sampling the $x$-direction. The solution yielded sharper discontinuities and a better $\mathcal{B}^x$ conservation, but still not sufficiently to perfectly resolve the large density jump, which requires an even higher resolution. 

\subsubsection{Loop advection}\label{sec:LoopAdvection}
We now test our code against the loop advection problem. This test is particularly challenging for grid based code due to the necessity to continuously evaluate strong fluxes, resulting from the fluid advection \citep{GardinerStone2005}. The Lagrangian nature of \texttt{GIZMO}, on the other hand, allows to naturally follow the uniformly moving fluid, reducing the effective fluxes between particles to the minimum.
For this setup, we initialize $2.5\times 10^5$ particles on a Cartesian grid, filling a periodic two-dimensional box of unit length. The fluid is given a uniform density ($\rho=1$) and specific internal energy ($\epsilon=4.5$), with $\Gamma=5/3$. The magnetic field is set to zero everywhere in the domain, but for a central circular region with radius $R_{\rm loop}=0.3$, in which we set an azimuthal magnetic field loop of the form
\begin{equation}\label{eq:LoopAdvection}
    (B^x,B^y,B^z) = 
    \begin{cases}
    A_{\rm loop} (-y/r,x/r,0); \ \  &r<R_{loop} \\   
    (0,0,0);  &r\geq R_{loop}
    \end{cases}
\end{equation}
where $A_{\rm loop}=10^{-3}$ and $r=\sqrt{x^2+y^2}$ is the cylindrical radius. Finally, we assign a bulk velocity $(v^x,v^y,v^z)=(1/2,1/12,0)$ to the fluid. We let the system evolve until $t=24$ in the MFM mode.

\begin{figure*}
\begin{center} 
\hspace{0.38cm}
\includegraphics[width=.37\textwidth]{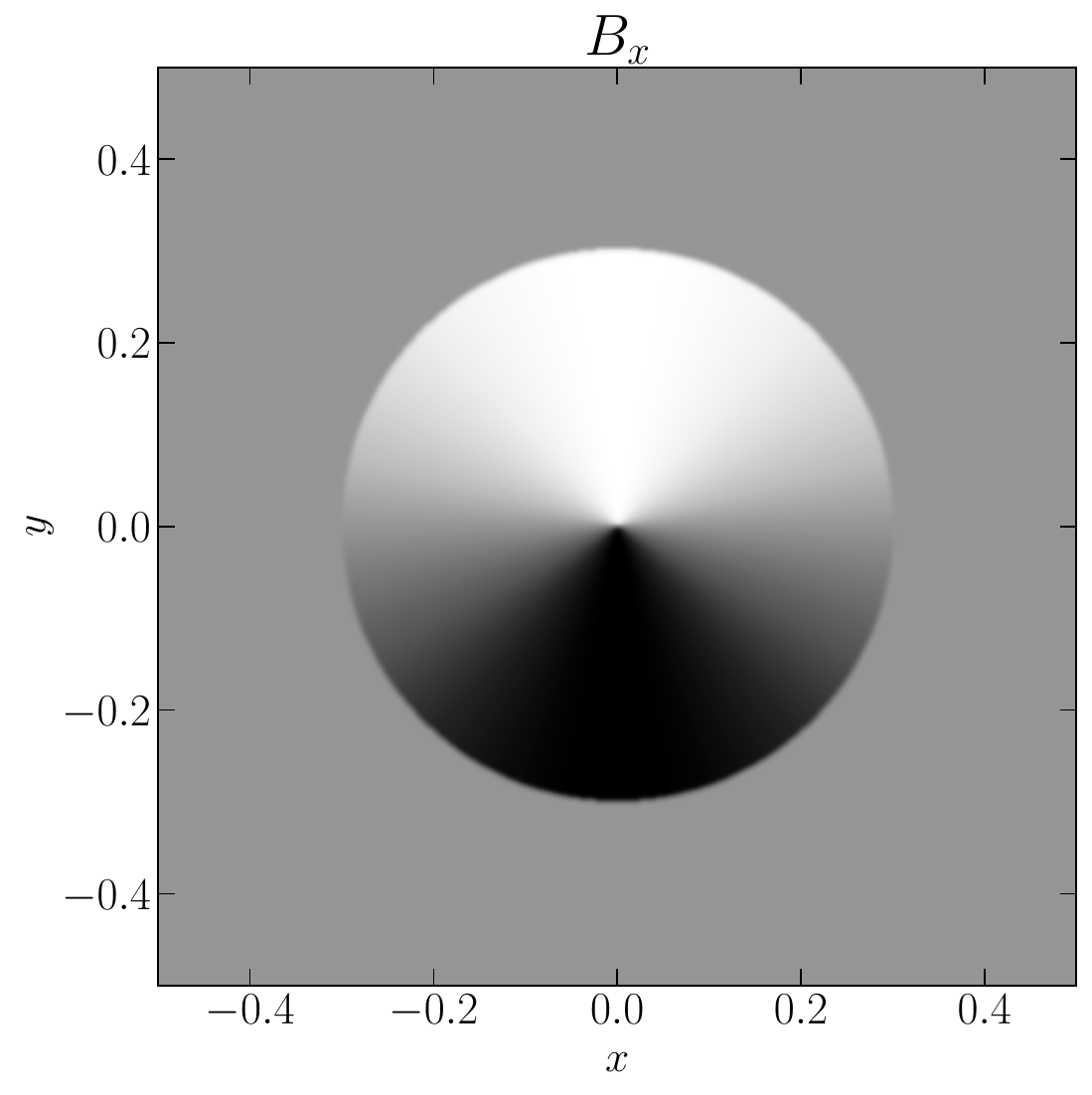}
\includegraphics[width=.45\textwidth]{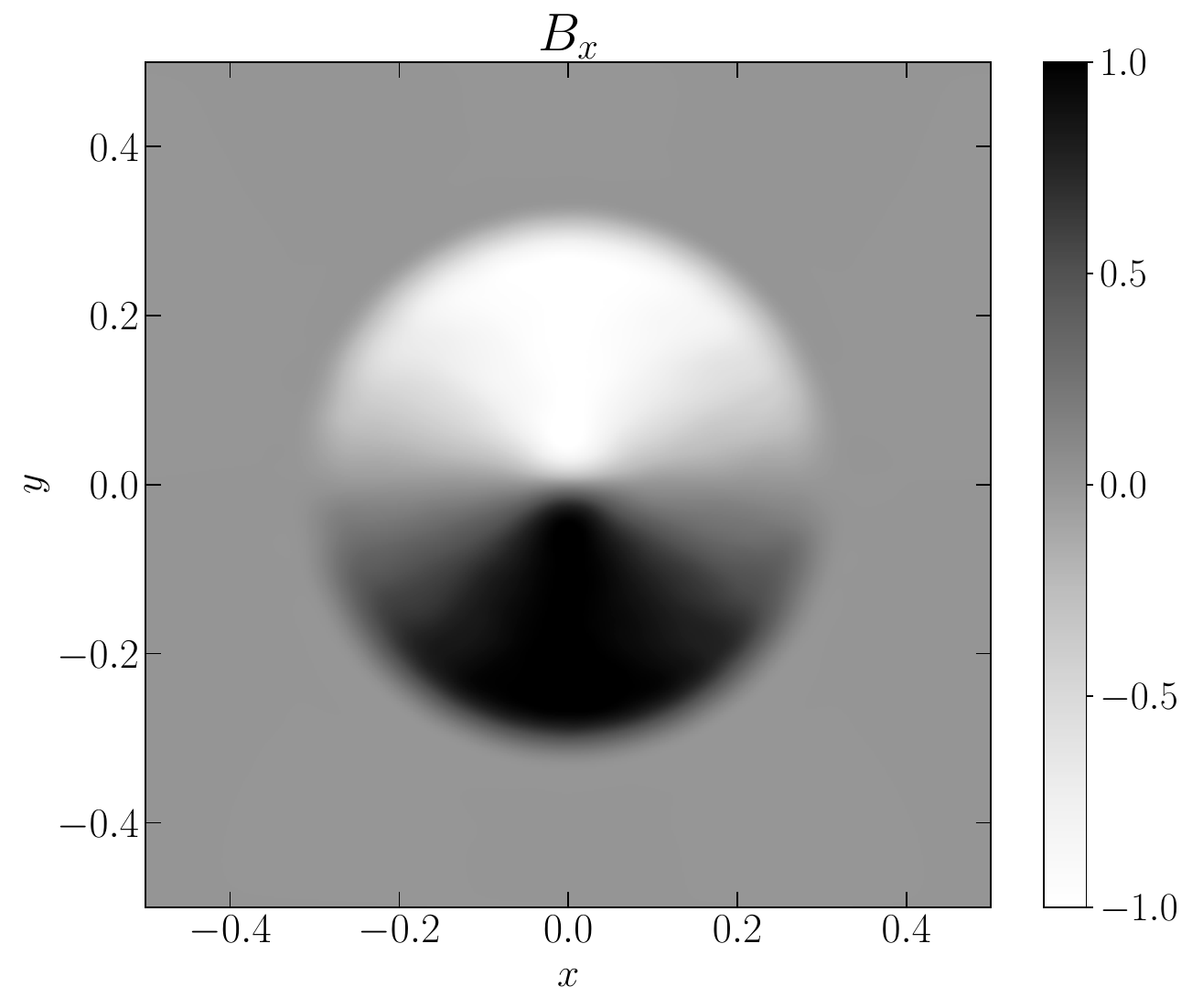}
\includegraphics[width=.37\textwidth]{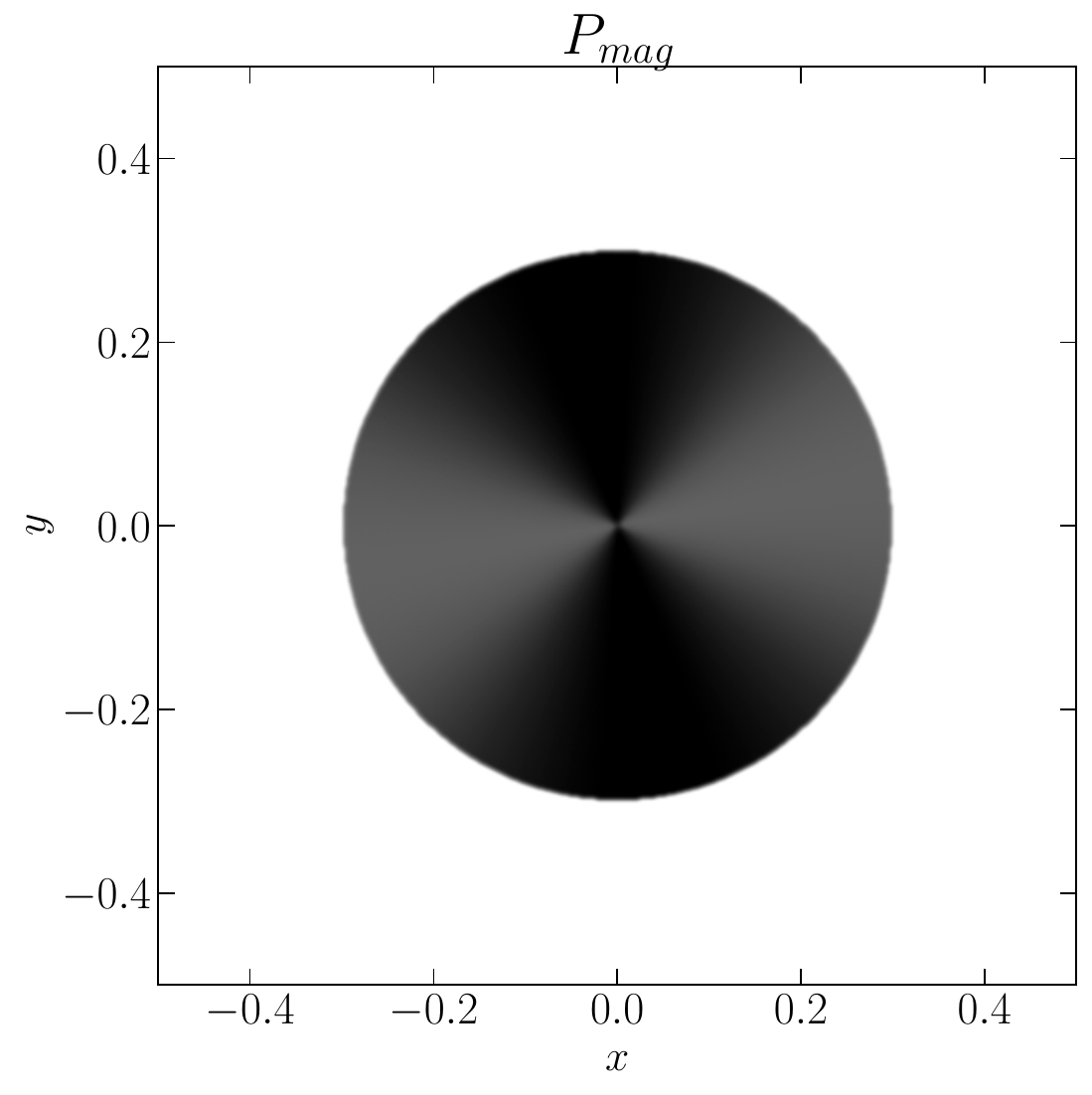}
\includegraphics[width=.43\textwidth]{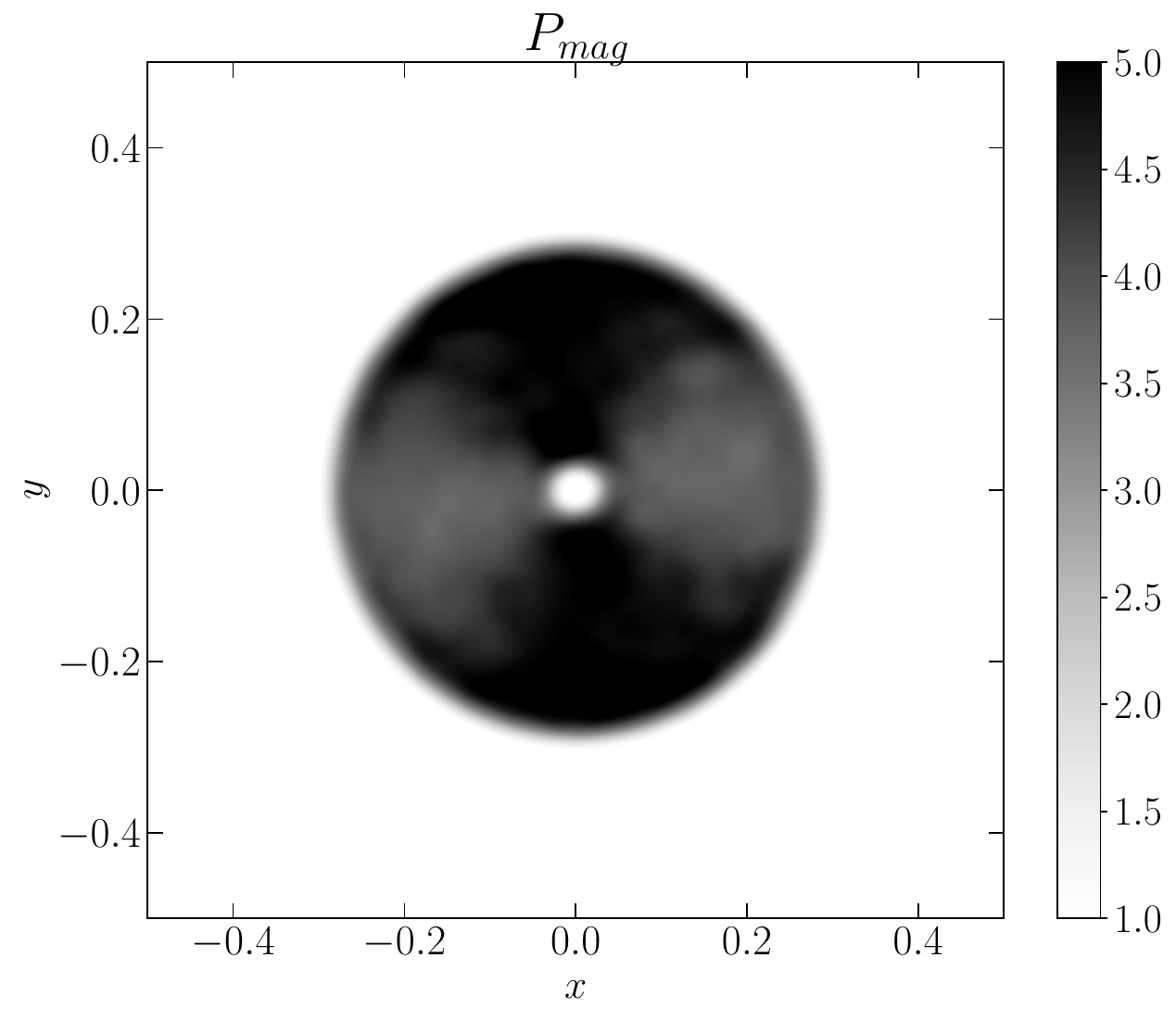}
\end{center}
\caption{Magnetic field loop advection performed in MFM mode. The top panels display the magnetic field intensity along the $x$ direction multiplied by $10^3$, whereas the bottom panels show the magnetic pressure $P_{mag}=b^2/2$ multiplied by $10^7$. Snapshots are taken at the initial time (left column) and at time $t=24$ (right column).}\label{fig:LoopAdvection}
\end{figure*}

In Fig. \ref{fig:LoopAdvection}, we plot $B^x$ and the magnetic pressure $P_{mag}=b^2/2$ at the initial and final times. Despite the presence of some smearing, due to the diffusive nature of our HLL solver, the magnetic field intensity and shape are well conserved throughout the advection. As also observed in many other codes \citep[]{Toth1996,Stone2008,Pakmor2011,Fragile2018,Fambri2018,Cipolletta2020} a magnetic pressure drop in a small central region is present. We note that, during this test, our scheme keeps the magnetic field divergence well controlled, reaching maximum values of $h(\partial_i\mathcal{B}^i)$ at $\sim 1\%$ the magnetic field intensity.

\subsubsection{Magnetic rotor}\label{sec:MagneticRotor}
As a second multidimensional stress test, we perform the relativistic version of the magnetic rotor \citep{DelZanna2003}. A 2D disk of radius $r_d=0.1$ and density $\rho_d=10$ is initially rotating with a uniform angular velocity $\Omega_d=9.95$. The rest of the domain is filled with a fluid of uniform density $\rho_{ext}=1$, initially at rest. The gas pressure is $P=1$ everywhere ($\Gamma=5/3$) and a magnetic field aligned with the $x$-direction of magnitude $B^x=1$ permeates the entire simulated box. This setup usually represents a really difficult test for numerical schemes, due to the high Lorentz factor $W\simeq10$, the formation of low density regions in the center of the rotor and the presence of strong discontinuities in the initial condition. To slightly alleviate this last point, we apply a smoothed transition on the rest mass density and on the velocity radial profiles between $r_{d}$ and $r_{ext}=0.115$ of the form $f(r)=(r_{ext}-r)/(r_{ext}-r_{d})$ so that
\begin{equation}\label{eq:MagneticRotorRho}
    \rho(r) =
    \begin{cases}
        \rho_d; \ \ \ &r\leq r_d \\
        1+9f(r); &r_d<r<r_{ext}\\
        \rho_0; &r\geq r_{ext}
    \end{cases}
\end{equation}
and 
\begin{equation}\label{eq:MagneticRotorV}
    |\textbf{v}(r)| =
    \begin{cases}
        r\Omega_d ; \ \ \ &r\leq r_d \\
        r_d\Omega_d f(r); &r_d<r<r_{ext}\\
        0; &r\geq r_{ext}.
    \end{cases}
\end{equation}
From these, we can compute the Lorentz factor $W(r)$ and the conserved density $D(r)$ radial profiles and we integrate the cumulative mass function $M(R<r)$. For the MFM run, we first place $\sim1.2\times  10^5$ particles on a closely packed lattice and we then stretch their radial position to match the cumulative mass function $M(R<r)$ up to $r=r_{ext}$, assuming equal-mass particles. Finally, outside of radius $r_{ext}$ we initialize a close packed lattice distribution of particles with uniform density $\rho_{ext}$. For the MFV case, we do not perform the radial stretch but we assign the particle mass to match $D(r)=\rho(r)W(r)$ instead, yielding an equal-volume particles distribution.

\begin{figure*}
\begin{center} 
\includegraphics[width=.4\textwidth]{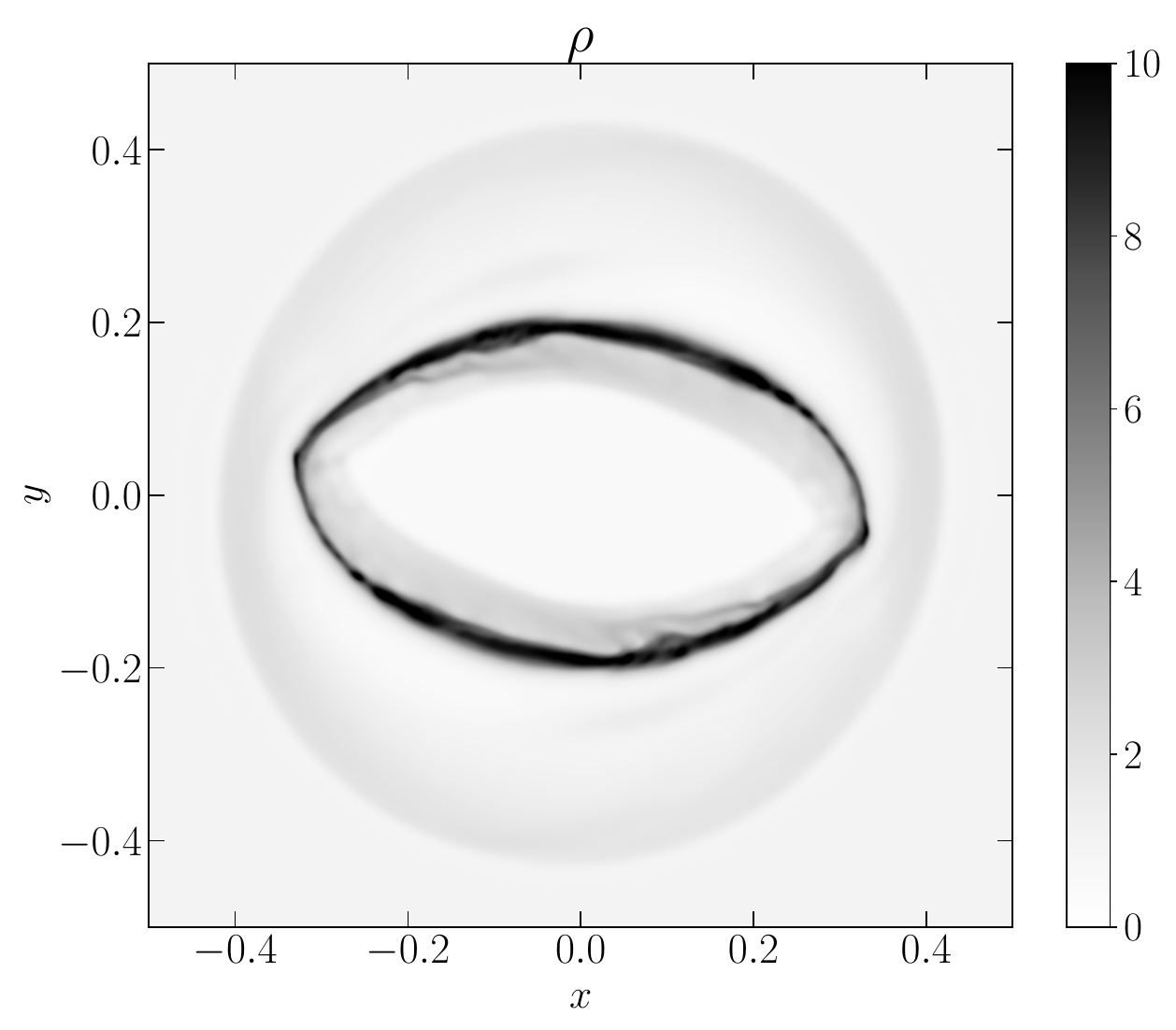}
\includegraphics[width=.4\textwidth]{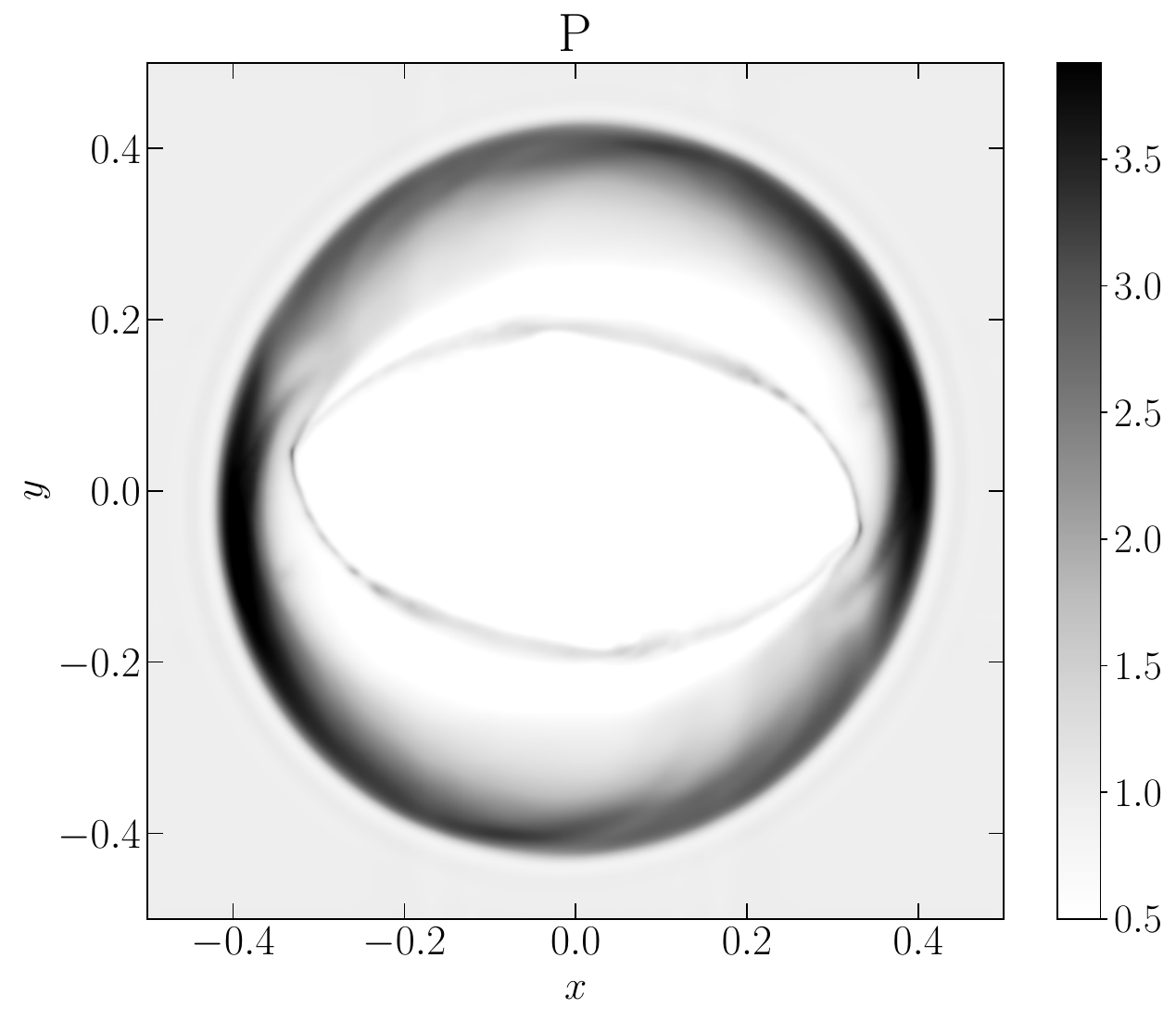}
\includegraphics[width=.4\textwidth]{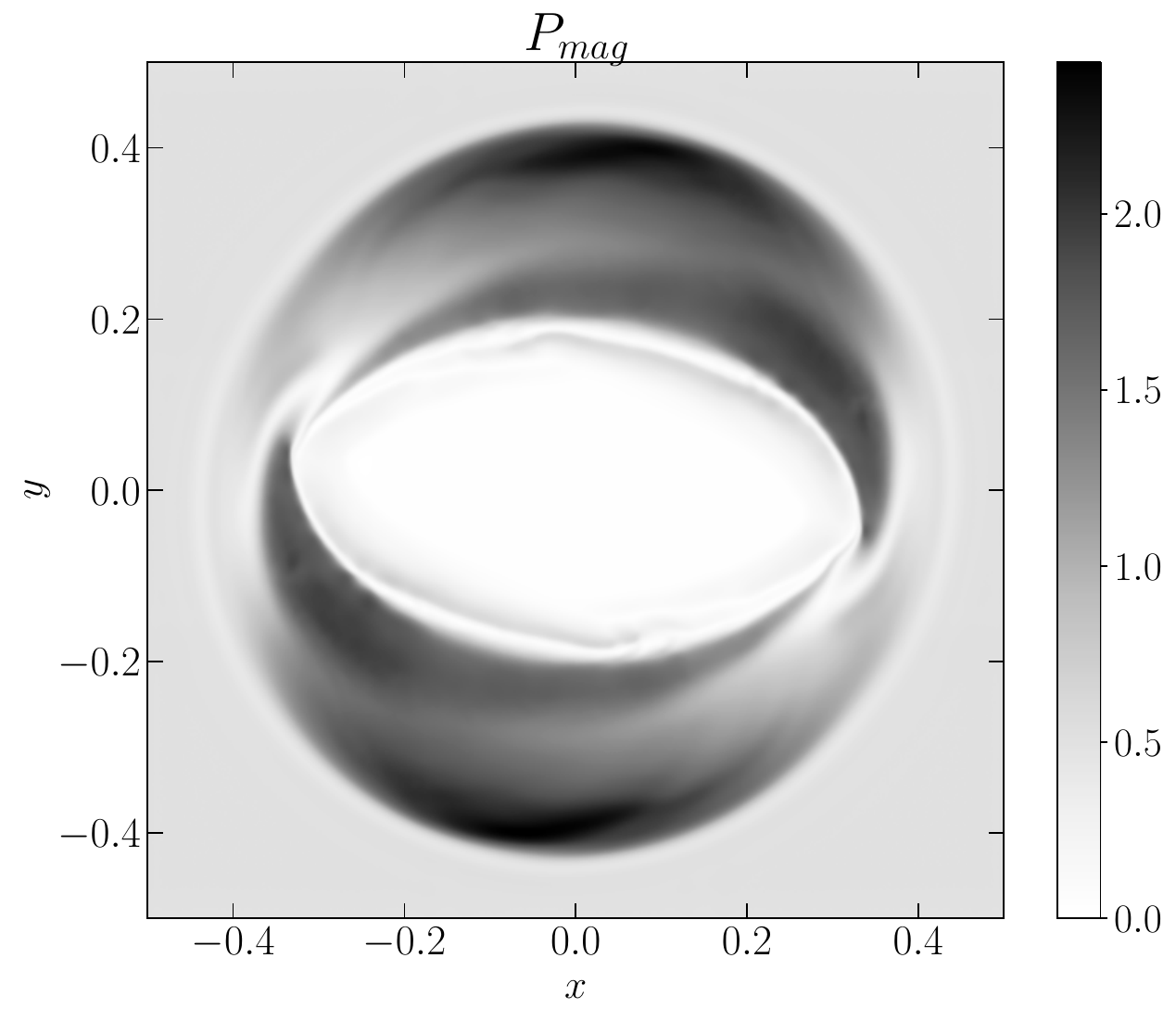}
\includegraphics[width=.4\textwidth]{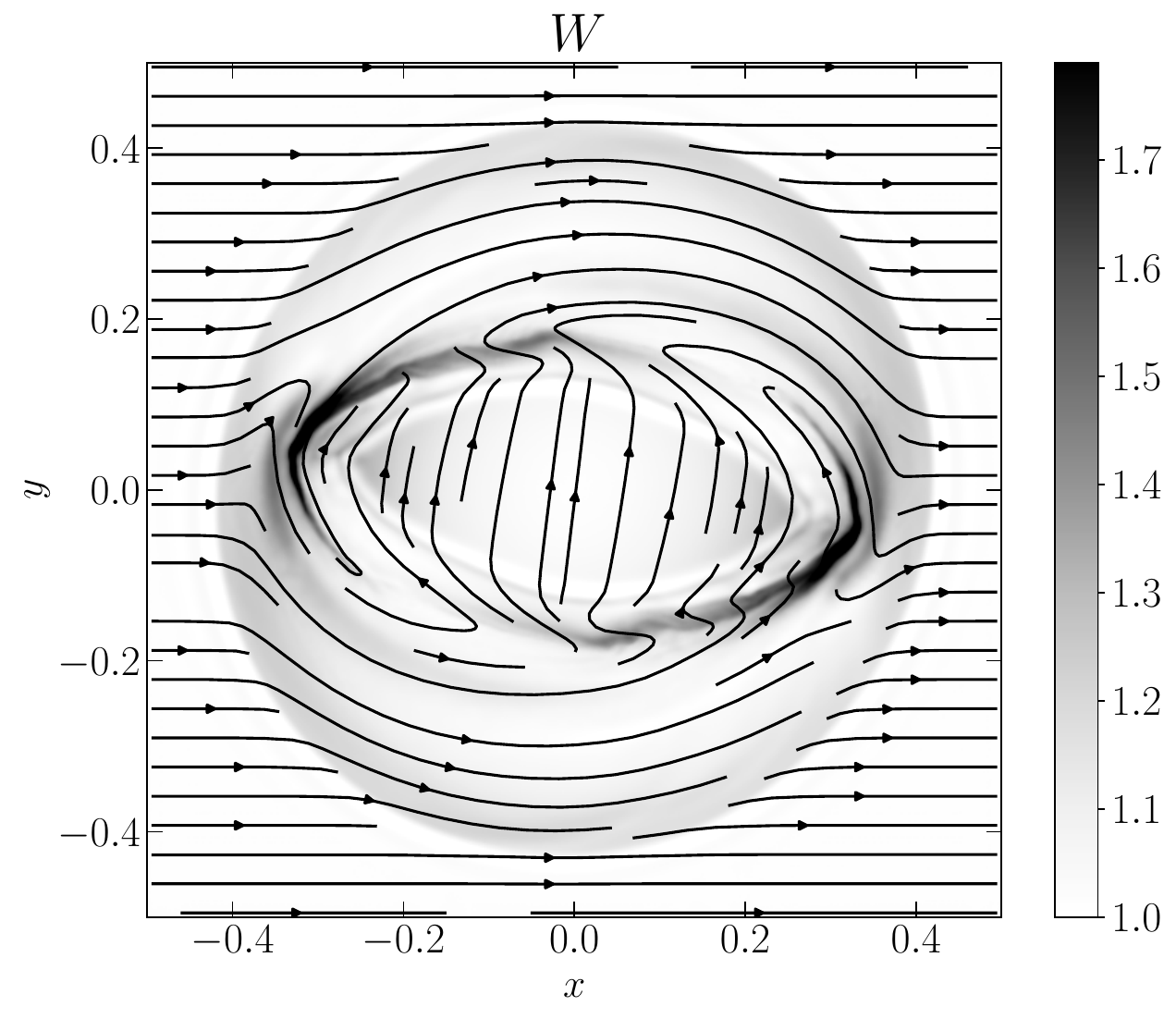}
\end{center}
\caption{The relativistic magnetic rotor problem at time t=0.4, performed in MFM mode. We plot 2D maps for the rest mass density (top-left), the gas pressure (top-right), the magnetic pressure (bottom-left) and the Lorentz factor with magnetic field lines (bottom-right).}\label{fig:MagneticRotor}
\end{figure*}

\begin{figure*}
\begin{center} 
\includegraphics[width=.9\textwidth]{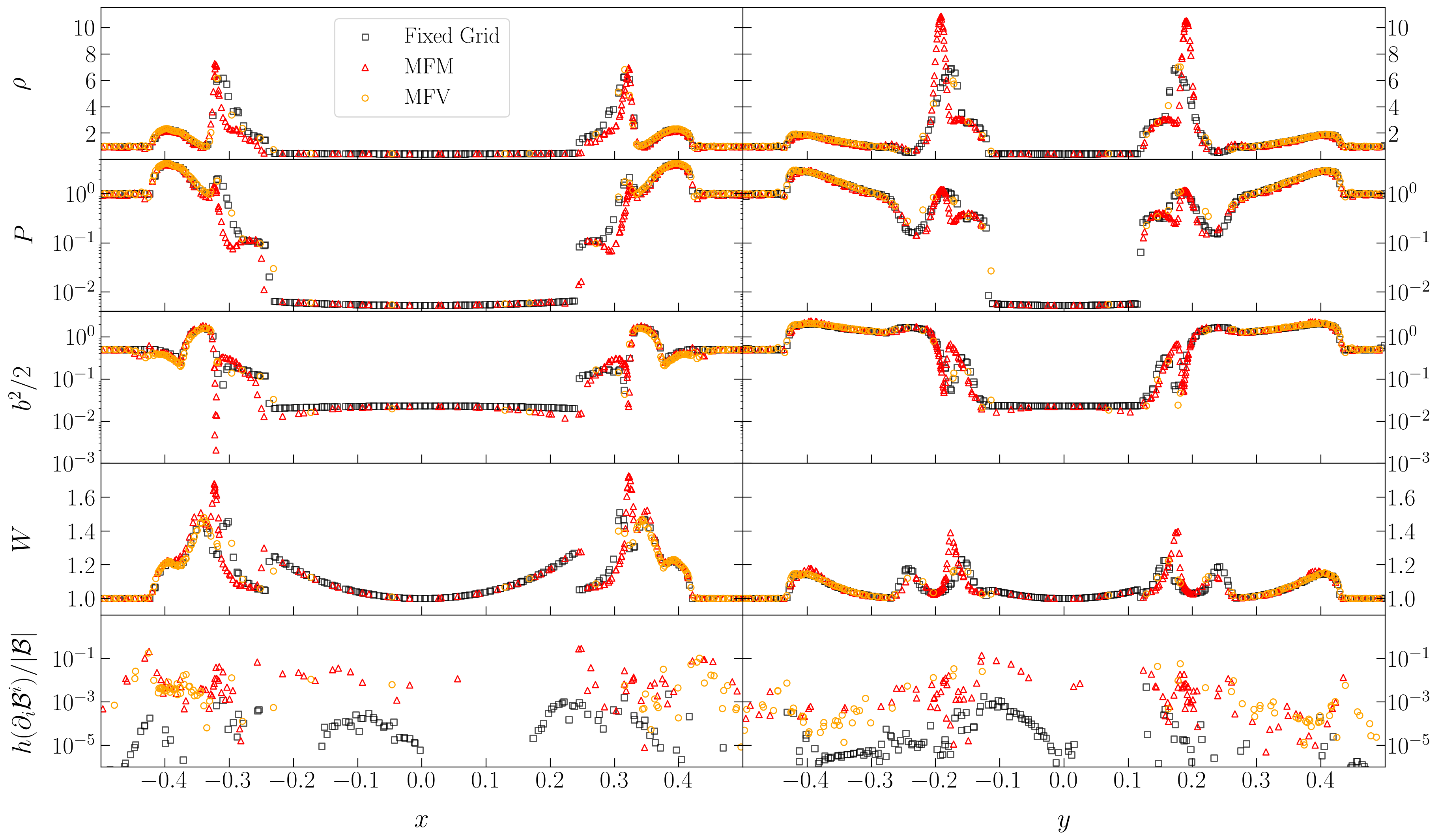}
\end{center}
\caption{Magnetic rotor test 1D slices along $y=0.5$ (left panels) and $x=0.5$ (right panels) of (from top to bottom) rest mass density, gas pressure, magnetic pressure, Lorentz factor and magnetic field divergence normalized to the magnetic field intensity, at time $t=0.4$. We show results from the MFM run (red triangles), MFV (orange circles) and from a fixed particle positions one (black squares) for reference.}\label{fig:MagneticRotorSlices}
\end{figure*}

Our code can evolve the system without breaking in both MFM and MFV modes. We show our results at time $t$=0.4 for the MFM case in Fig. \ref{fig:MagneticRotor}. Our solution exhibits good behavior overall: the center of the disk is quickly emptied of gas, while an overdensity propagates outside. The presence of the magnetic field slows down the rotation of the fluid, yielding a maximum Lorentz factor of $W\simeq2.48$ at the final time; due to the flux-freeze effect, magnetic field lines are twisted by the fluid in the central region. When compared to results from grid based codes \cite[e.g.][]{DelZanna2003,DuffellMacFadyen2011,GRHydro}, our evolution scheme yields higher maximum values of the rest mass density and of the Lorentz factor, peaked on the oblate shear front. This behavior is better quantified in Fig. \ref{fig:MagneticRotorSlices}, where we plot the most important quantities along horizontal and vertical slices, at the output time $t=0.4$. For a more comprehensive view, we display here the results from the MFV run too, alongside the ones computed with fixed particle positions for reference. While our MFV results show a good agreement with the reference run and with the ones presented in \citet{GRHydro}, the MFM run yields higher rest mass density and Lorentz factor peaks by a factor $\sim1.5$ and $\sim1.3$ respectively; these features produce a slightly faster expansion of the rotor. We note that the resolution in regions with higher densities in our MFM run is $\sim2$ times higher than the MFV one (and than the resolution used in the referenced grid based codes), which could therefore better resolve the compression of the fluid.\footnote{We tested this hypothesis by performing a high resolution ($4\times$) run using the MFV method and obtaining a rest mass density peak of $\rho_{max}>9$ and a maximum Lorentz factor $W>1.8$. These values exceed those of the low-resolution reference case by factors of approximately 1.3 and 1.25, respectively.}

\subsubsection{Cylindrical blast wave}\label{sec:CylindricalBlastWave}
One of the most demanding special relativistic tests for a multidimensional code is the blast wave explosion. We test our scheme against the magnetized cylindrical case presented in \citet{Komissarov1999} and later performed as a test for various codes \citep[]{DelZanna2007,BeckwithStone2011,GRHydro,Cipolletta2020}.
In this problem an initial overdense and overpressurised gas of $\rho_{in}=10^{-2}$ and $P_{\rm in}=1$ is confined inside a radius $r_{in}=0.8$. The external medium is instead uniform with rest mass density $\rho_{ext}=10^{-4}$ and pressure $P_{ext}=3\times 10^{-5}$. Between the inner $r_{in}$ and an external radius $r_{ext}=1$, we apply a transition on the rest mass density of the form
\begin{equation}\label{eq:ICtransition}
    \ln\rho(r)=\ln\rho_{in} + \left( \frac{r_{in}-r}{r_{in}-r_{ext}}\right)\ln\left(\frac{\rho_{ext}}{\rho_{in}}\right)
\end{equation}
and an equivalent one on the pressure profile. A magnetic field $\textbf{B}=(0.1,0,0)$ permeates the entire domain and the adiabatic index is set to $\Gamma=4/3$.

As done for the magnetic rotor test (Sec. \ref{sec:MagneticRotor}), for the MFM run we initialize $\sim1.6\times 10^5$ equal-mass particles in a closely packed lattice and we then stretch their radial position to match the integrated mass cumulative function $M(R<r)$ up to $r=r_{ext}$. Outside this radius, we do not perform any stretching, initializing the fluid with a uniform density $\rho_{ext}$. For the MFV run, we place $\sim2.6\times 10^6$ particles on a closely packed lattice and assign the particle mass to match the $\rho(r)$ profile.\footnote{The higher particles number is needed to obtain a resolution on the blast wave front similar to the MFM run.} We let the system evolve until $t=3$ in a periodic box of side length $6$, using a Dedner damping parameter $K=0.75$. Our code is able to evolve this difficult test in both MFV and MFM mode, with and without the Dedner divergence cleaning scheme.\footnote{In the MFV case, the velocity of each particle is `smoothed' with the velocities of the interacting neighbors, assuming a contribution of 20\% instead of the standard 30\%.}

\begin{figure*}
\begin{center} 
\includegraphics[width=.4\textwidth]{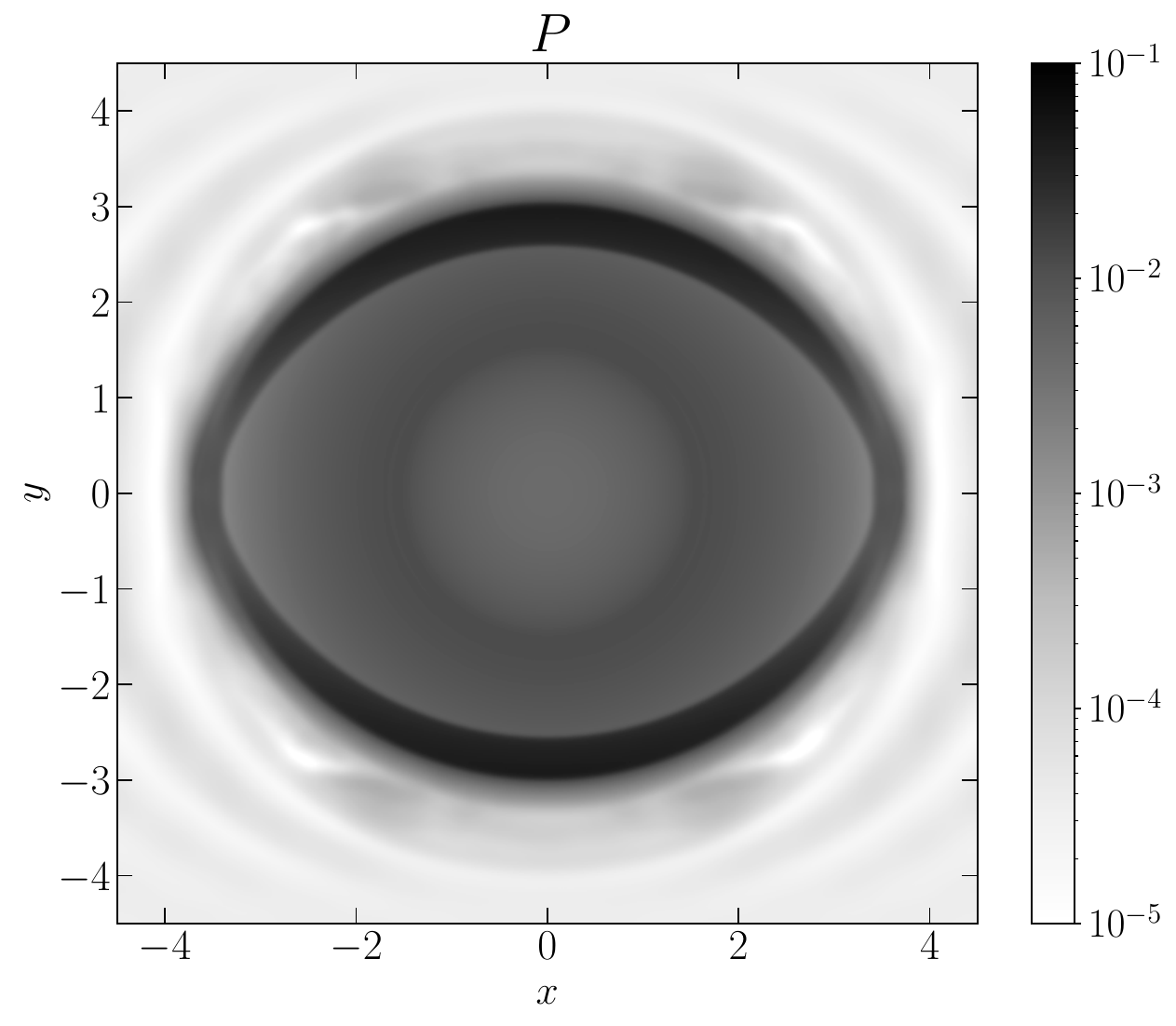}
\includegraphics[width=.4\textwidth]{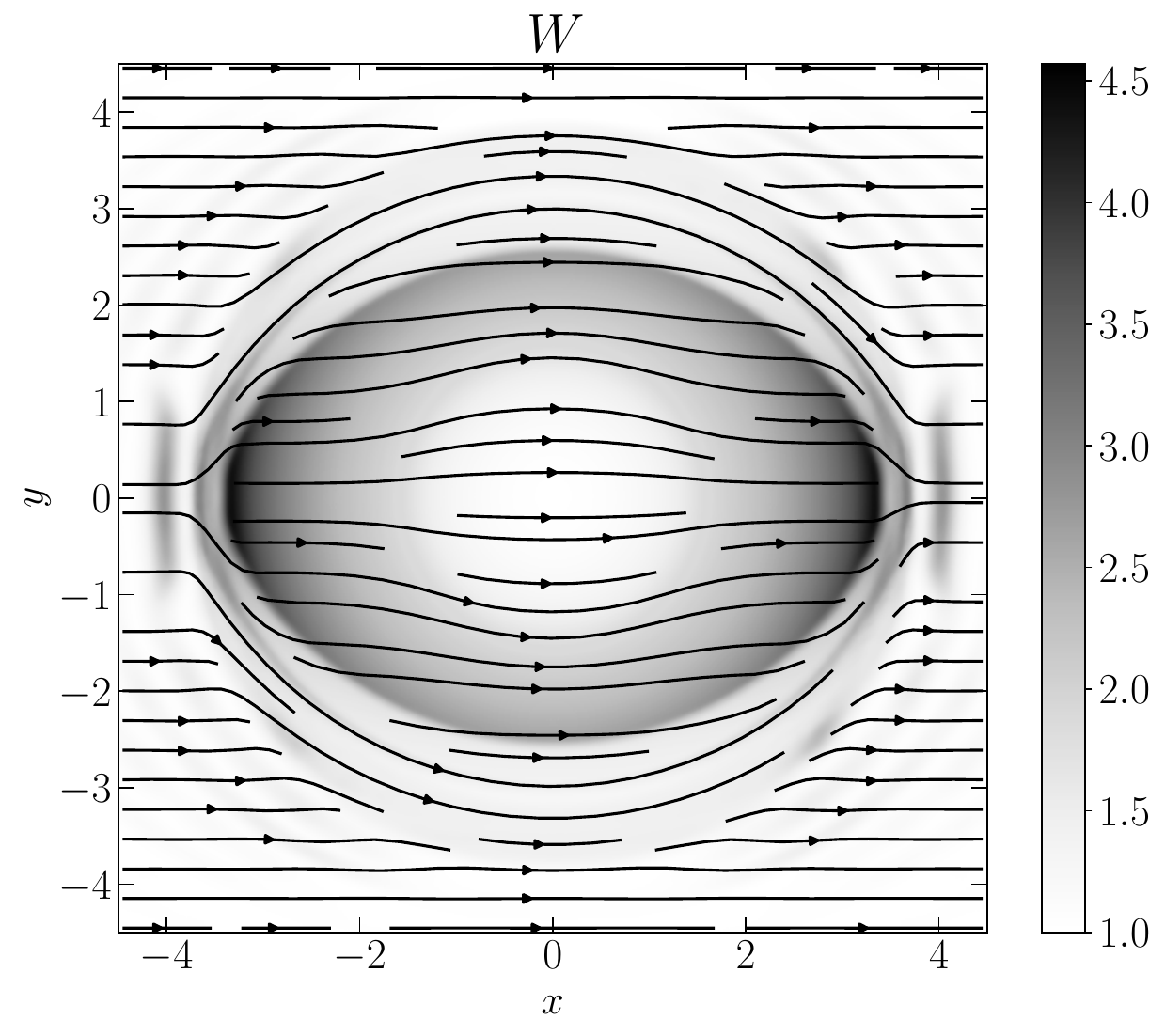}
\includegraphics[width=.4\textwidth]{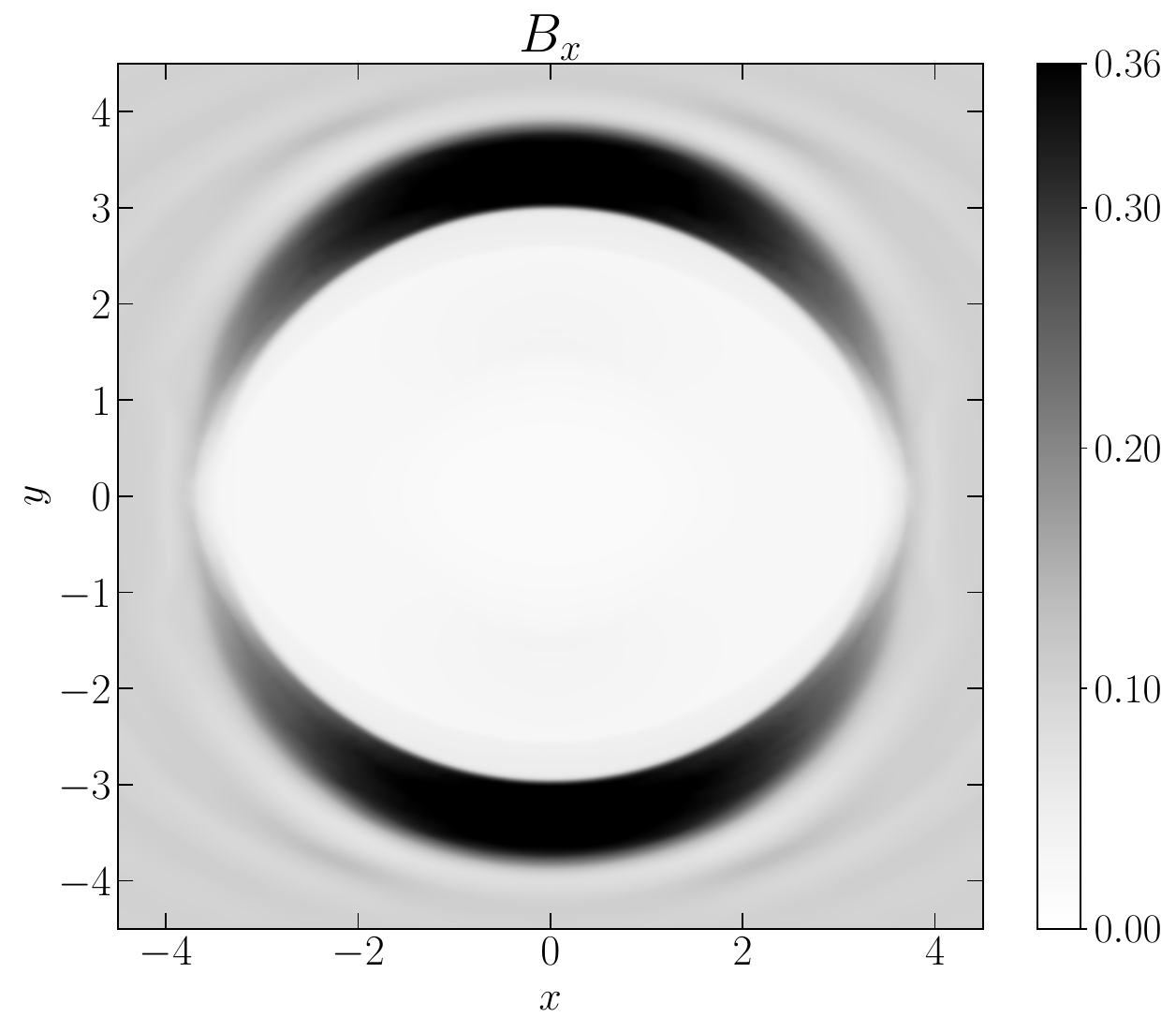}
\hspace{0.1cm}\includegraphics[width=.41\textwidth]{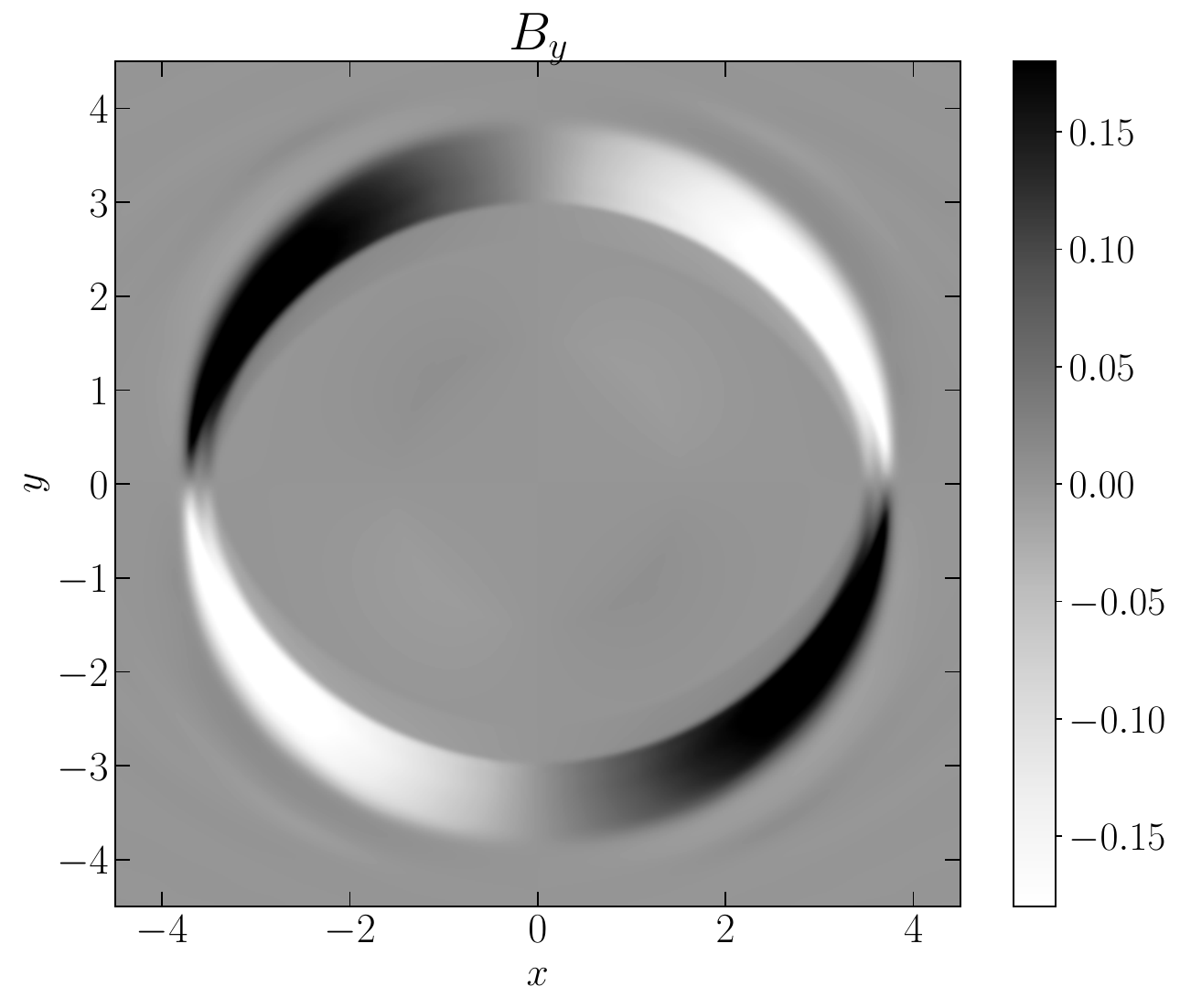}
\end{center}
\caption{The cylindrical blast wave problem at time $t=3$, performed in MFM mode. We plot 2D maps for the rest mass density (top-left), the gas pressure (top-right), the magnetic field x-component (bottom-left) and its y-component (bottom-right).}\label{fig:CylindricalBlast}
\end{figure*}

\begin{figure*}
\begin{center} 
\includegraphics[width=.3\textwidth]{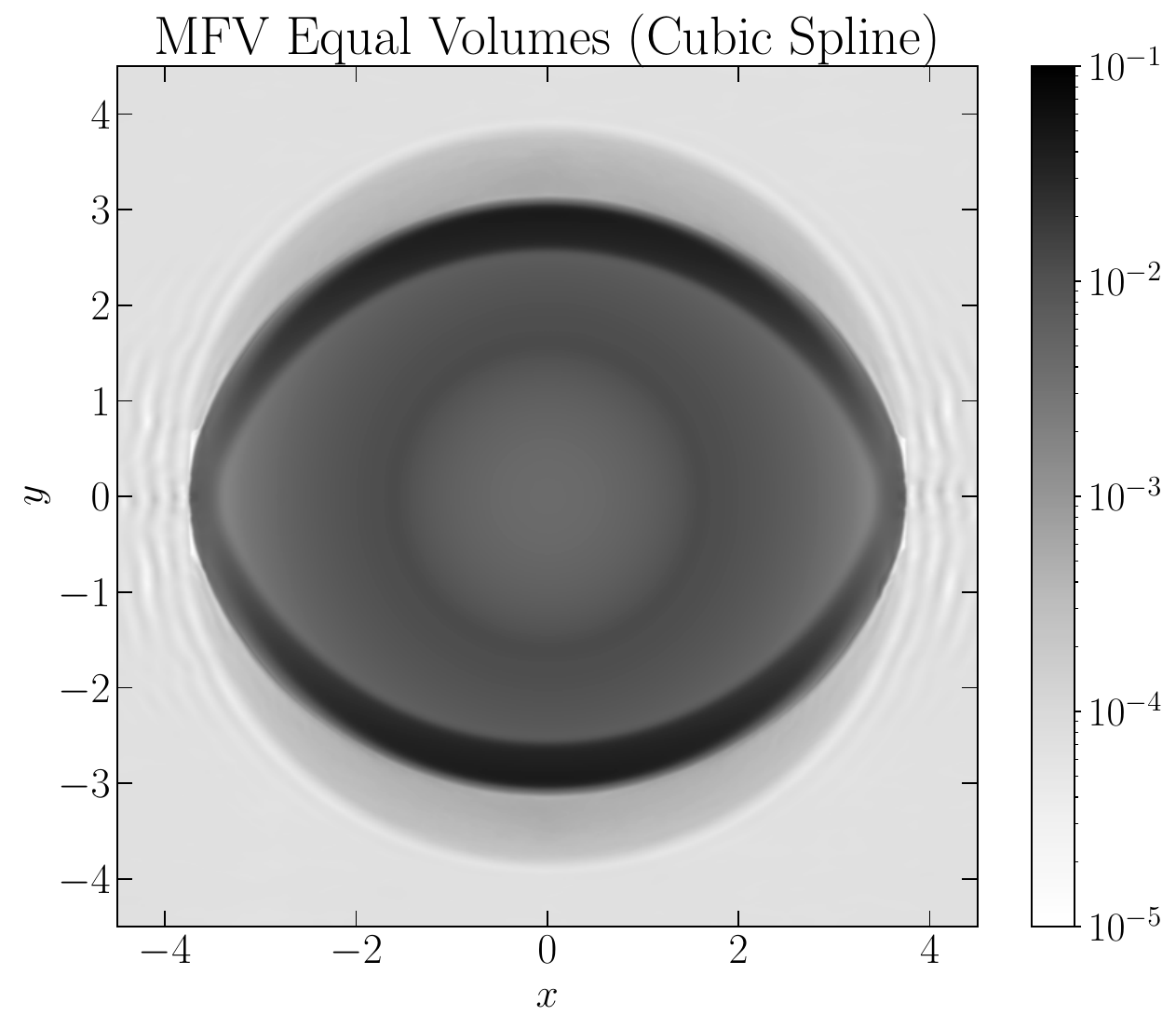}
\includegraphics[width=.3\textwidth]{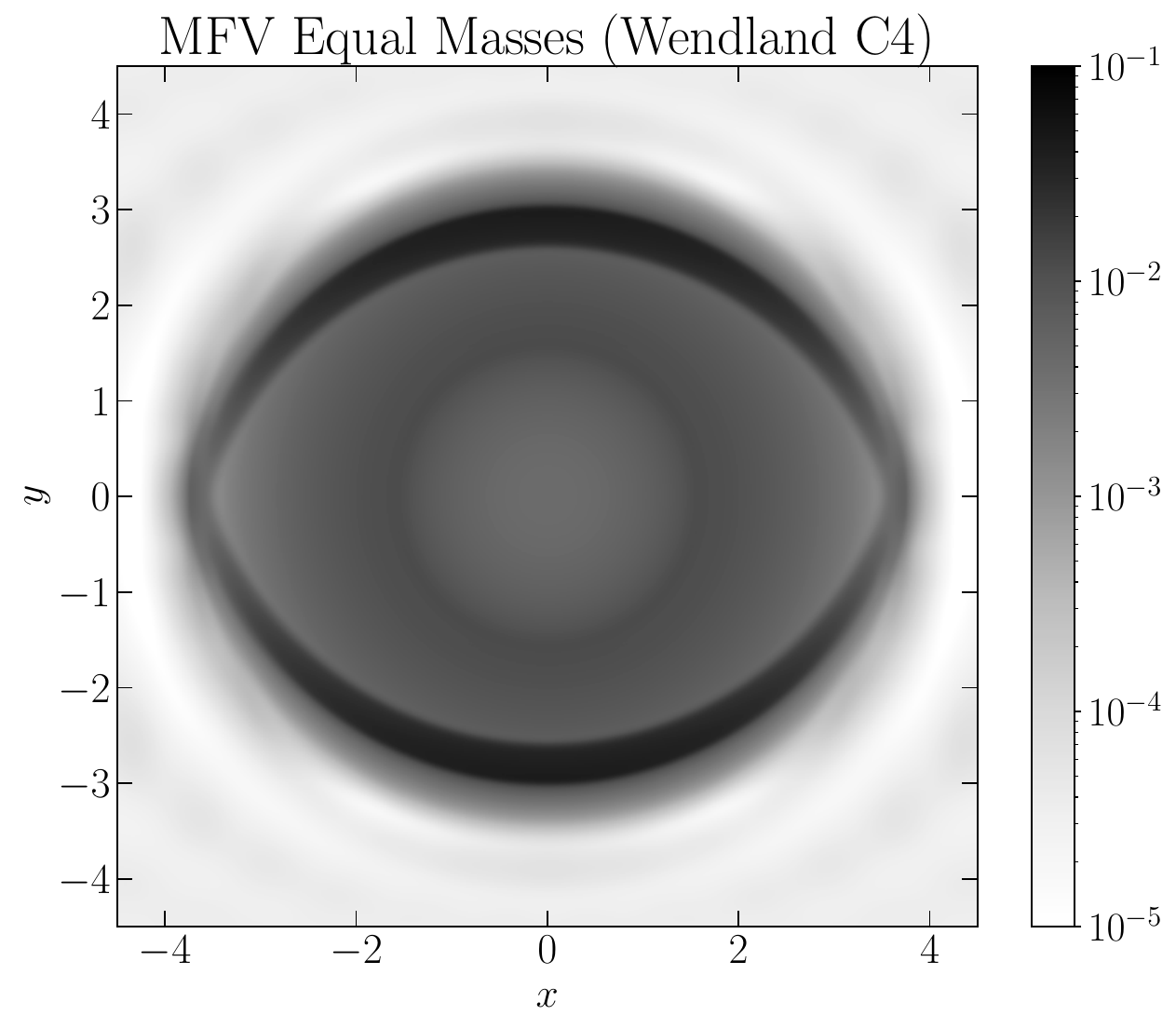}
\includegraphics[width=.3\textwidth]{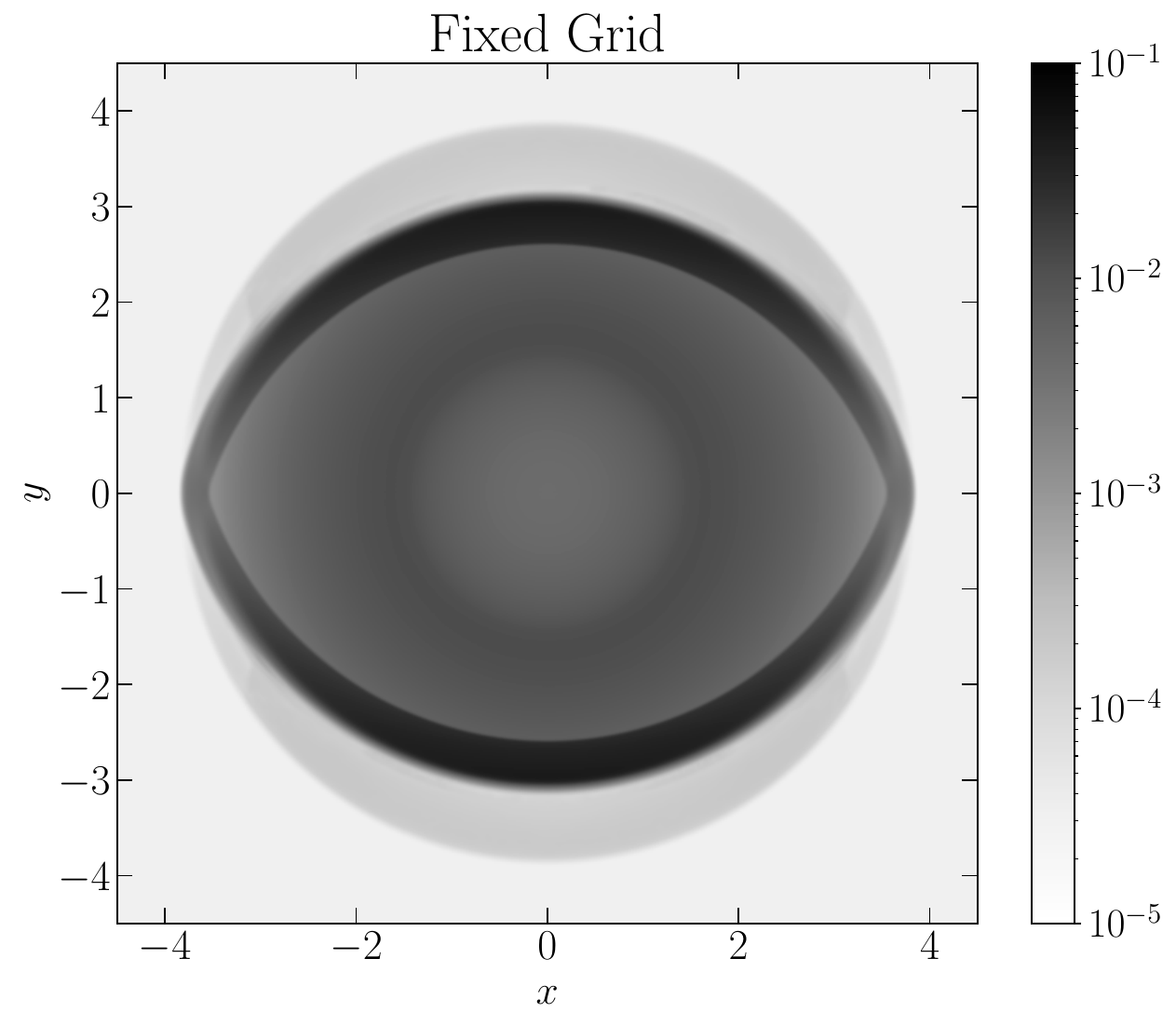}
\end{center}
\caption{Pressure distributions in the blast wave problem for three different run: (left) MFV mode, equal-volume particles, using a cubic spline as kernel function; (center) MFV mode, equal-mass particles, using a Wendland C4 as kernel function; (right) fixed particle positions run.}\label{fig:CylindricalBlastExtra}
\end{figure*}

We plot the MFM run results in Fig. \ref{fig:CylindricalBlast}. Our solution exhibits some oscillations in the direction aligned with initial magnetic field; their intensity is initially more relevant on the velocity evolution but later affects the other quantities. This is the same numerical artifact discussed in Sec. \ref{sec:Shocks} for the Balsara1 test, caused by the strong volume compression of particles ahead of the shock front. Again, this effect can be attenuated employing lower-order reconstruction\footnote{Note that the default slope limiters in \textsc{gizmo} have larger tolerance on the reconstruction compared to those commonly employed in other codes.} of the magnetic field or a larger kernel. In Fig. \ref{fig:CylindricalBlastExtra}, we display the pressure distribution resulting from three additional runs. Our MFV run shows fewer oscillations outside the blast wave, due to the smaller volume gradient on the shock in the initial condition, but with a small background pressure enhancement. The equal-mass particle configuration evolved in the MFV mode and using a larger kernel \citep[Wendland C4, ][]{Wendland1995,DehnenAly2012} yields slightly smaller oscillations in amplitude but with more diffusion. Finally, we show our result evolved with fixed particle positions (as in a fixed grid setup); this last plot is comparable with results from other grid-based codes \citep[]{GRHydro,DelZanna2003}.

\begin{figure*}
\begin{center} 
\includegraphics[width=.9\textwidth]{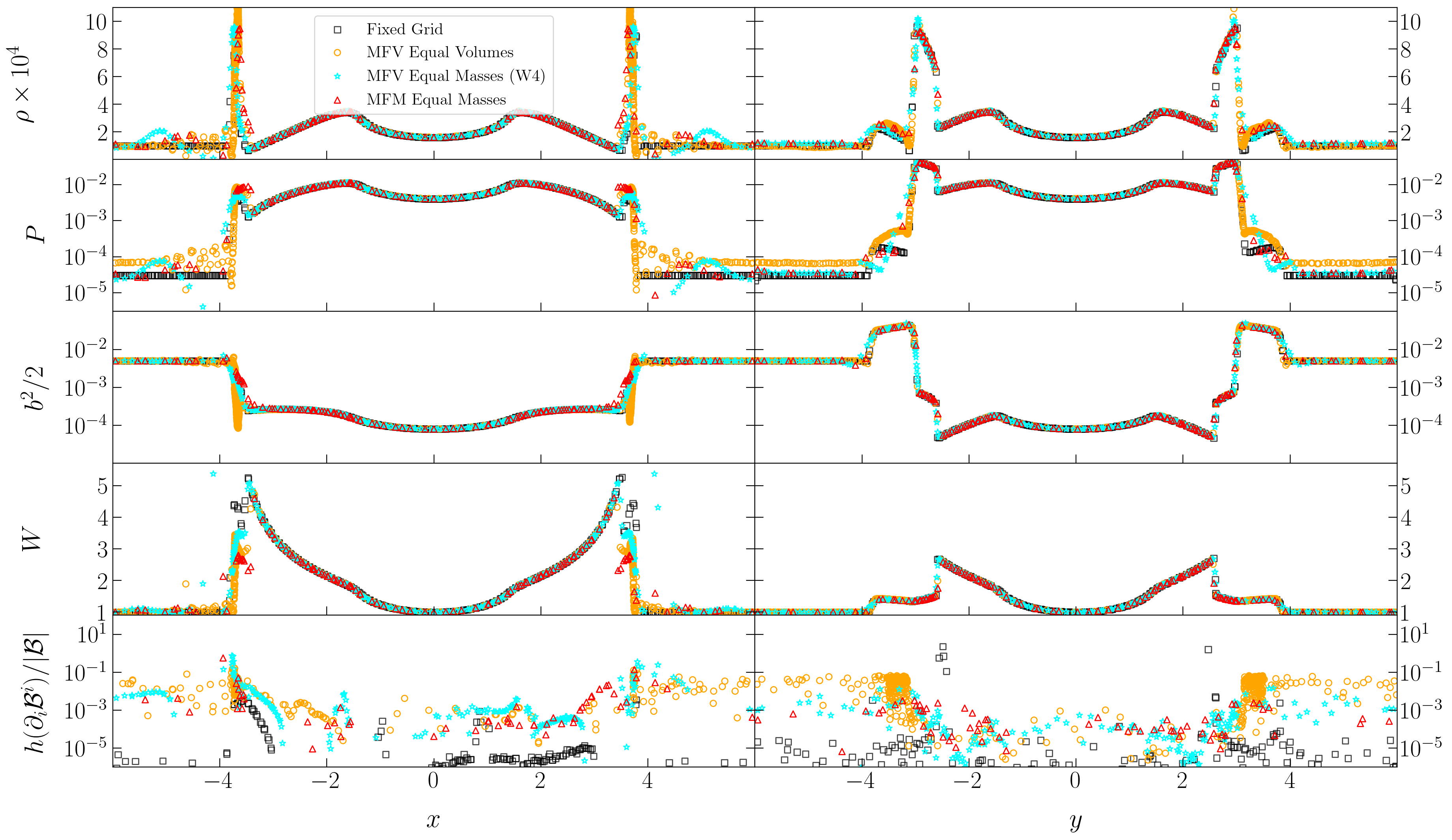}
\end{center}
\caption{Cylindrical blast wave test 1D slices along $y=0$ (left panels) and $x=0$ (right panels) of (from top to bottom) rest mass density, gas pressure, magnetic pressure, Lorentz factor and magnetic field divergence in units of the magnetic field intensity, at time $t=3$. We show results from the MFM run (red triangles), MFV with initially equal volumes particles (orange circles), MFV with initially equal masses particles evolved using the Wendland C4 kernel (cyan stars) and from a fixed-grid one (black squares) for reference.}\label{fig:CylindricalBlastSlices}
\end{figure*}

In Fig. \ref{fig:CylindricalBlastSlices}, we plot the results evaluated along horizontal and vertical slices. We can recognize an small drop in the magnetic pressure in the MFV equal-volumes run in the direction aligned with the initial magnetic field. The MFM run slightly underestimates the velocity of the shock in the $x$-direction, yielding a lower Lorentz factor. The configuration evolved using the Wendland C4 kernel seems to produce the best solution overall (more similar to the fixed grid case).

\subsubsection{Spherical blast wave}\label{sec:SphericalBlastWave}
As a final special relativistic test, we perform the spherical blast wave, an extension of the 2D cylindrical blast wave discussed in Sec.~\ref{sec:CylindricalBlastWave}, another very challenging problem for GRMHD codes. We initialize the system as in \citet{Cipolletta2020}, i.e. an inner region ($r\leq r_{in}=0.8$) is filled with a fluid with density $\rho_{in}=10^{-2}$ and pressure $P_{in}=1$. At $r\geq r_{ext}=1$, the gas density is $\rho_{ext}=10^{-4}$ and the pressure $P_{ext}=3\times 10^{-5}$. A smooth transition links the two regions, as in the cylindrical case. The magnetic field, aligned with the $z$-direction, is set to $\textbf{B}=(0,0,0.1)$, and fills the whole domain. The adiabatic index is $\Gamma=4/3$ everywhere. We initialize $\sim 1.8\times 10^6$ equal-mass particles in a stretched closely packed lattice and we let the system evolve until $t=4$, with a Dedner damping parameter $K=0.75$.

\begin{figure*}
\begin{center} 
\includegraphics[width=.42\textwidth]{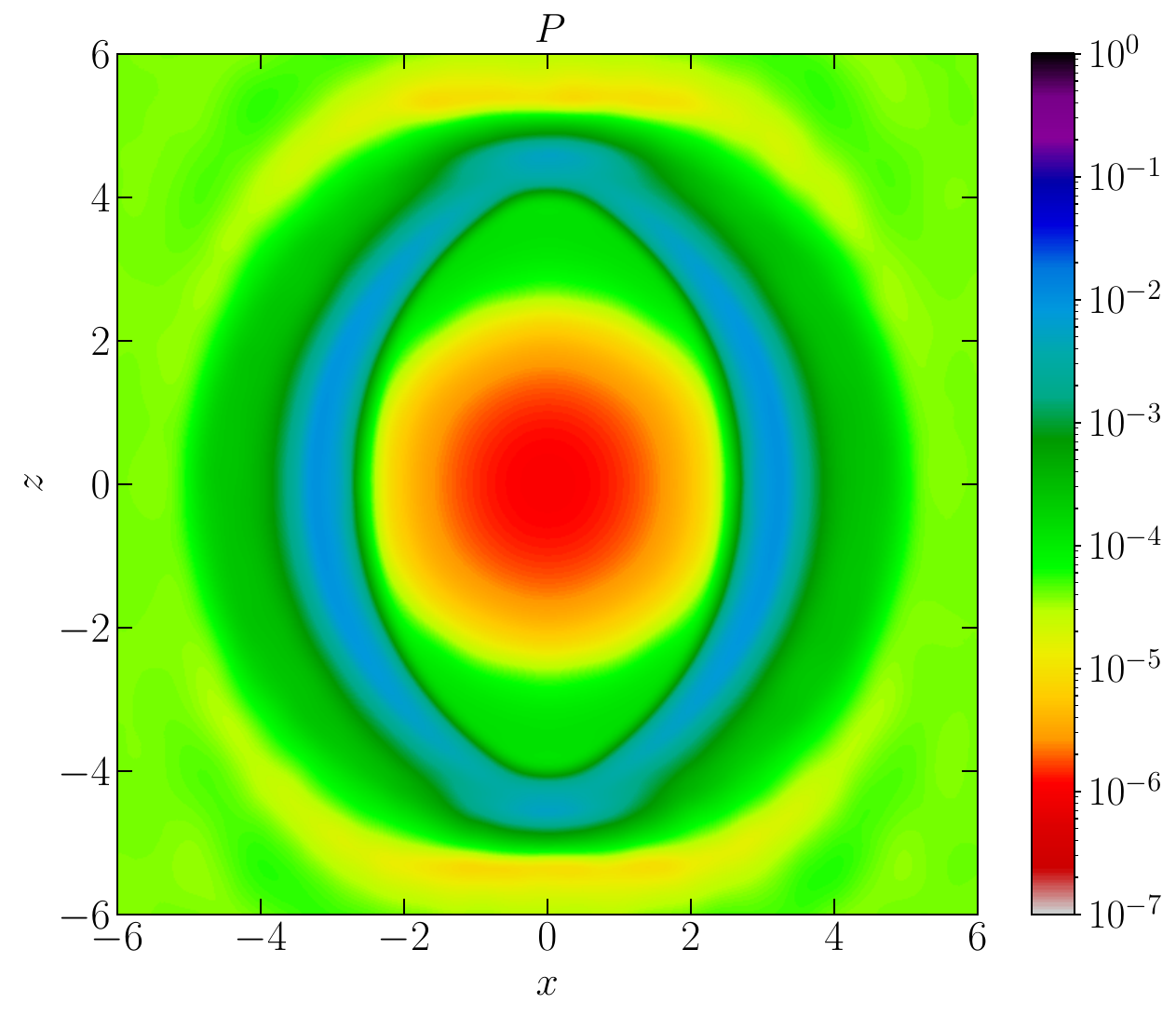}
\includegraphics[width=.4\textwidth]{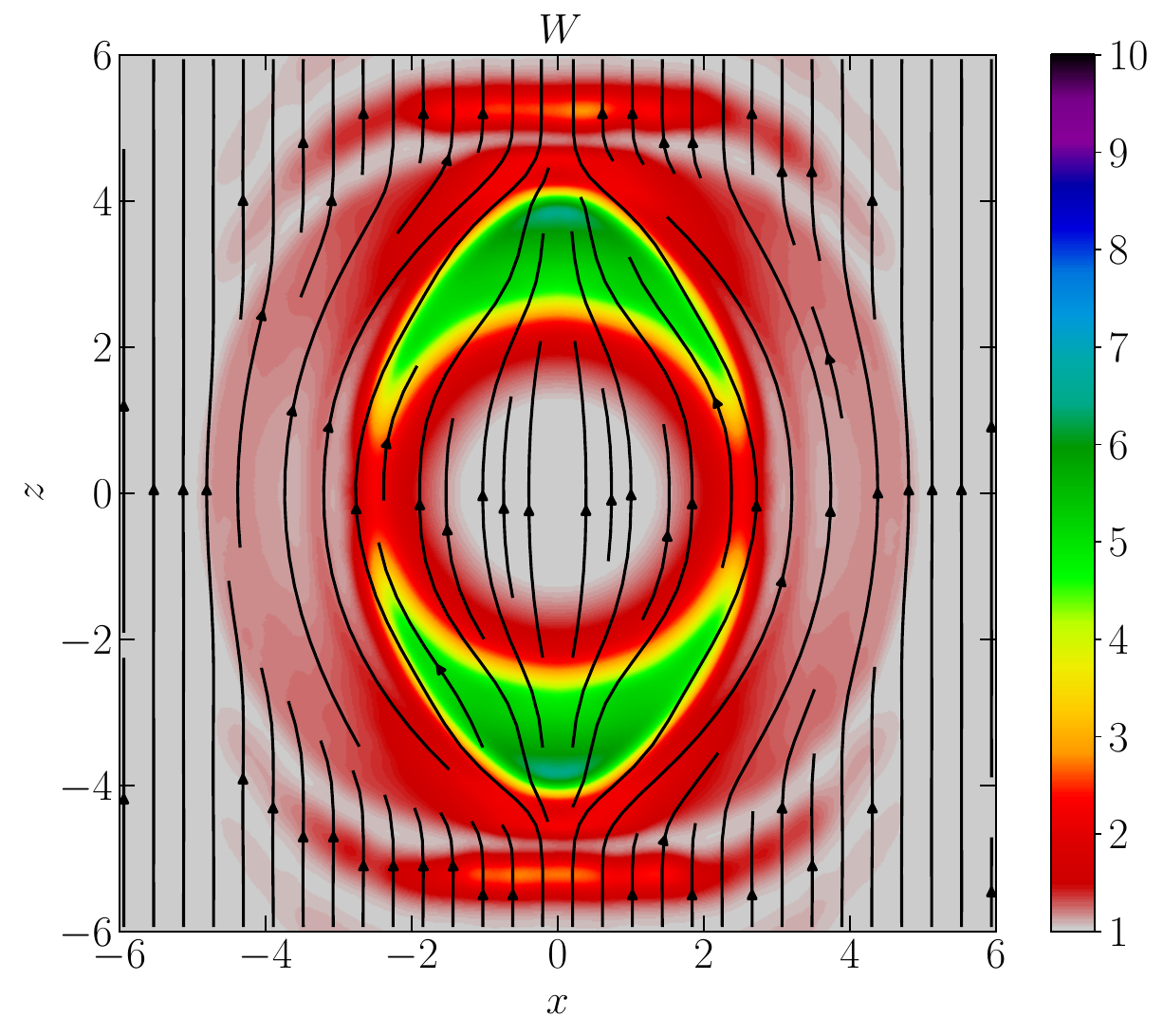}
\end{center}
\caption{Two dimensional slices at $t=4$ and $y=0$ of the spherical blast wave problem, evolved in MFM mode. We report the gas pressure in the left panel and the Lorentz factor with the magnetic field lines distribution overlaid in the right panel.}\label{fig:SphericalBlast}
\end{figure*}

Because of the intrinsically larger number of neighbors employed in 3D, both the `cell' volumes and the magnetic field divergence error (Eq. \ref{eq:DivBEvaluation}) estimates improve in accuracy. Nonetheless, some spurious oscillation in the Lorentz factor ahead of the shock are still observed, affecting the density and pressure evolution. Despite these artefacts, which we expect can be alleviated by suitably tuning the kernel shape, extension, and the slope limited employed (as already discussed), the code is able to evolve this challenging test without breaking and without significantly ruining the solution within the shock front. Moreover, we do not observe any grid alignment effect or diagonal artifacts (usually found in other GRMHD codes \citep{DelZanna2003}).

\subsection{General relativistic magnetohydrodynamics}\label{sec:GRTests}
In this section, we present two general relativistic MHD tests, that allow us to determine the robustness of our scheme in non-Minkowskian background spacetimes. 
\subsubsection{Magnetized TOV star}\label{sec:TOVStar}
\begin{figure}
\begin{center} 
\includegraphics[width=.45\textwidth]{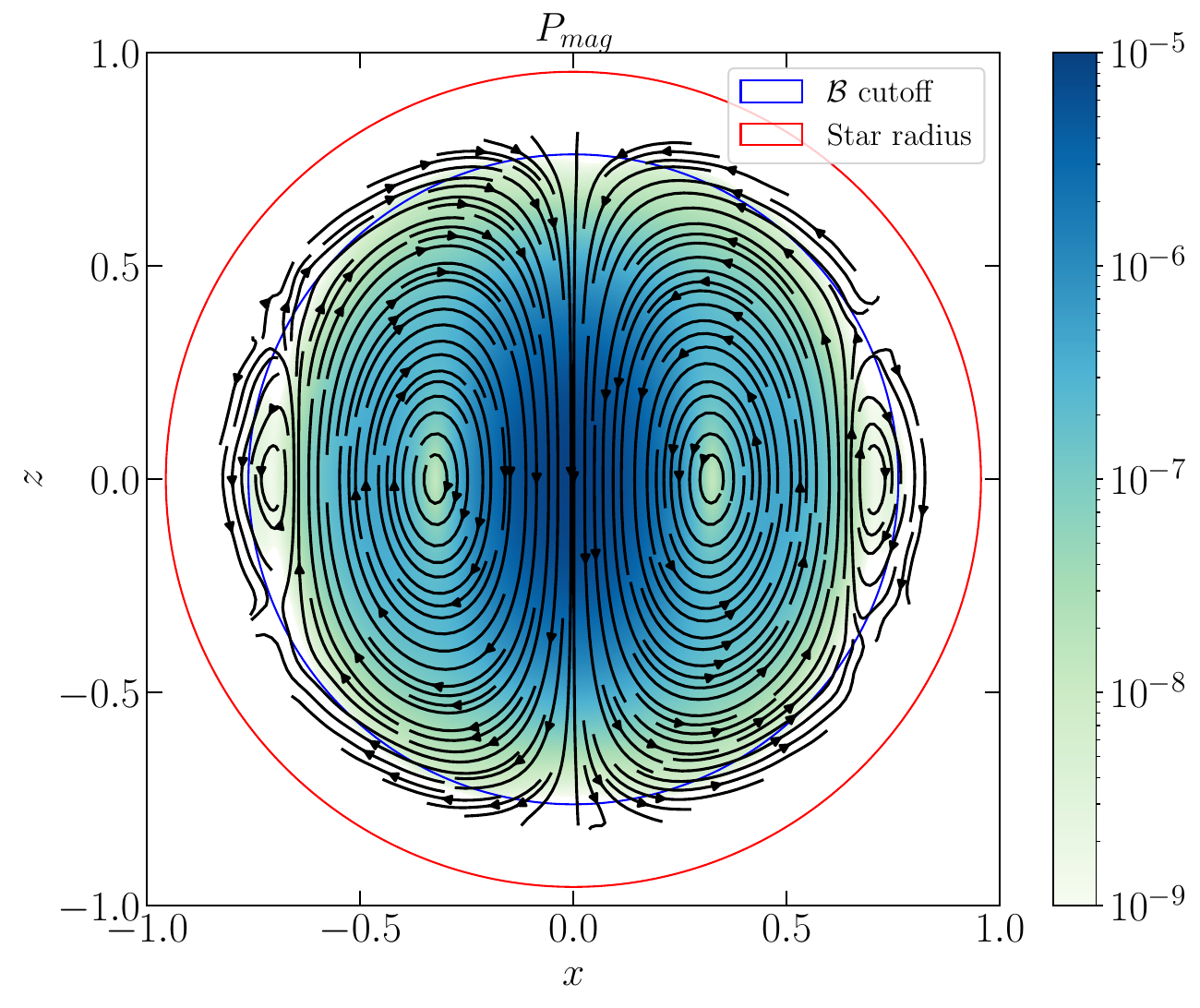}
\end{center}
\caption{Initial magnetic field configuration of the TOV test. We plot the magnetic pressure with superimposed magnetic field lines. The red circle marks the star radius (defined as the point where $P=10^{-8}P_{max}$), while the blue one indicates the radius at which the magnetic field goes to zero. The magnetic field is zero in white-colored regions.}\label{fig:TOVBField}
\end{figure}

As a first test, we consider the evolution of a magnetized, non-rotating Tolman-Oppenheimer-Volkoff (TOV) star \citep[]{TOV,TOV2}. With this problem, we can test the correct implementation of the terms proportional to the metric functions $\sqrt{\gamma}$ and $\alpha$ defined in the ADM formalism.
The initial magnetized setup is presented in many GRMHD codes \citep[]{Duez2006,liu2008,GRHydro,Cipolletta2020} and represents a perturbation of the pure GRHD configuration, already tested in the relativistic \texttt{GIZMO} paper \citep{Lupi2023}; hence, we use the same initial configuration with $\sim10^6$ particles, adding a poloidal magnetic field resulting from a purely toroidal vector potential $A_\phi$ given by
\begin{equation}\label{eq:TOVVectorPotential}
    A_\phi = A_b\omega^2\max(P-P_{cut},0)^{\xi},
\end{equation}
where $\omega=\sqrt{x^2+y^2}$ is the cylindrical radius. Here we employ $\xi=2$, which ensure that the pressure's derivative is continuous at all radii, $A_b=15$, to have a maximum magnetic-to-gas pressure parameter $\beta^{-1}\sim8\times 10^{-4}$, and $P_{cut}=0.04P_{max}$, that gives the radius at which the magnetic field goes to zero,  where $P_{max}$ is the maximum initial gas pressure (at the center of the star). We then project $A_\phi$ on the Cartesian directions as $A_i = \frac{\partial x^\phi}{\partial x^i}A_\phi =(-yA_\phi/\omega^2,xA_\phi/\omega^2,0)$ and compute the magnetic field as $\mathcal{B}^i=\sqrt{\gamma}n_\mu\varepsilon^{\mu ijk}\partial_jA_k =\sqrt{\gamma}\alpha\varepsilon^{ijk}\partial_jA_k$, obtaining
\begin{equation}\label{eq:TOVMagneticField}
    \mathcal{B}^i=\sqrt{\gamma}\alpha A_b
    \begin{cases}
        \frac{xy}{r}\partial_r\tilde{P};\\
        \frac{yz}{r}\partial_r\tilde{P};\\
        -(\frac{\omega^2}{r}\partial_r\tilde{P} + 2\tilde{P});
    \end{cases}
\end{equation}
where we have defined $\tilde{P}\equiv\max(P-P_{cut},0)^{\xi}$. We plot the resulting magnetic configuration (magnetic pressure distribution and magnetic field lines) on the $x-z$ plane in Fig. \ref{fig:TOVBField}. This prescription represents a small perturbation of the pure hydrodynamical equilibrium studied in \citet{Lupi2023}.

We evolve the TOV star using both MFM and MFV modes, with and without the Dedner cleaning scheme. We note that our solution is stable even when employing the Powell cleaning terms only, although we do not report the results here. In the fiducial case, that employs the complete divergence cleaning scheme, we employ $K=0.1$ and $f=2.5$; since we are evolving an equilibrium configuration with low magnetization, the estimated signal velocities are very slow, hence the need to overestimate the cleaning speed by a higher factor $f$. We note that evolving the system using $f=1$ does not corrupt the GRMHD solution and only produce slightly higher $\partial_i\mathcal{B}^i$ values.

\begin{figure*}
\begin{center} 
\includegraphics[width=.9\textwidth]{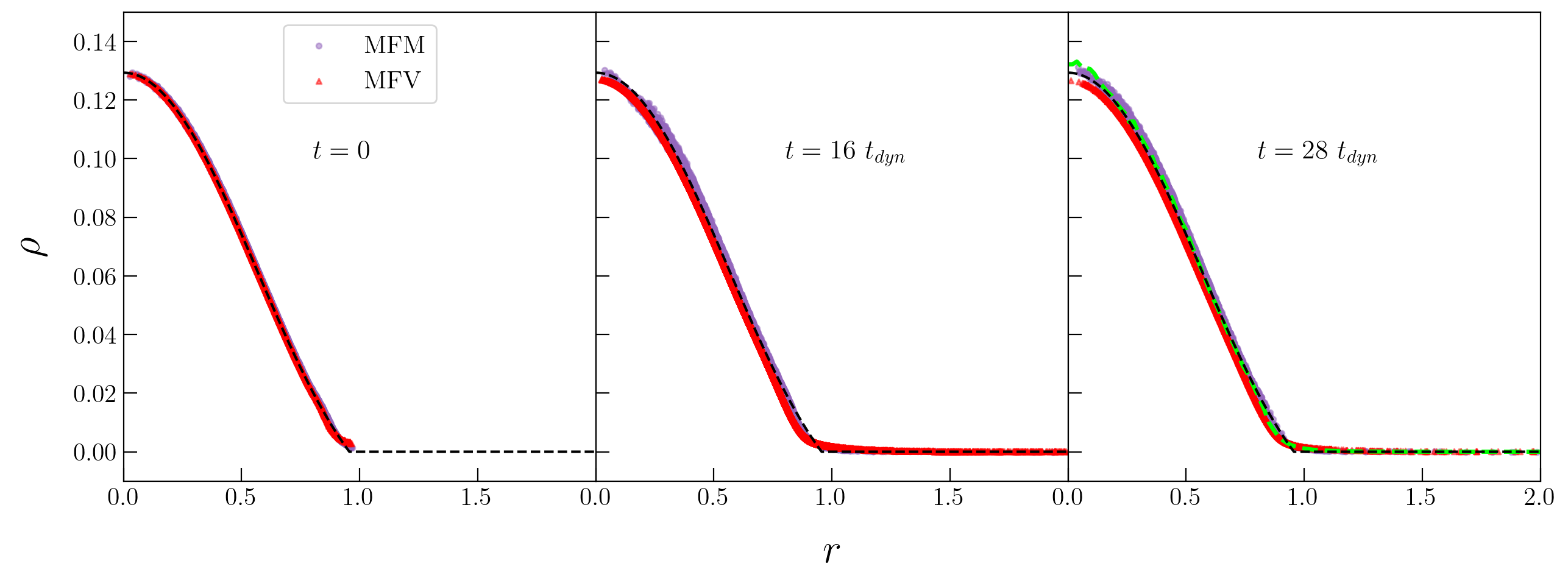}
\end{center}
\caption{Radial profile of TOV equilibrium at $t=0$, $t\approx16t_{dyn}$ and $t\approx28t_{dyn}$ for the MFM (purple circles) and the MFV (red triangles) run. The black dashed line marks the exact solution. The MFM solution at time $t\approx50t_{dyn}$ is displayed as a lime dashed line.}\label{fig:TOVRadialProfile}
\end{figure*}

We let the runs evolve for $t\simeq28 t_{dyn}$, where we have defined the dynamical time $t_{dyn}\equiv 1/\sqrt{\rho_c}$ (as in the GRHD test, the central rest mass density of the TOV star is $\rho_c\approx0.129285$). In Fig. \ref{fig:TOVRadialProfile} we show the radial rest mass density profile at three different snapshots, both for the MFV and MFM runs. Our results agree with the exact solution and we do not recognize big differences from the hydrodynamical case presented in \citet{Lupi2023}. At the end of the MFV run the central density has lowered by $\sim2\%$ from its initial value, while we witness a small increase in the MFM run.

\begin{figure}
\includegraphics[width=.5\textwidth]{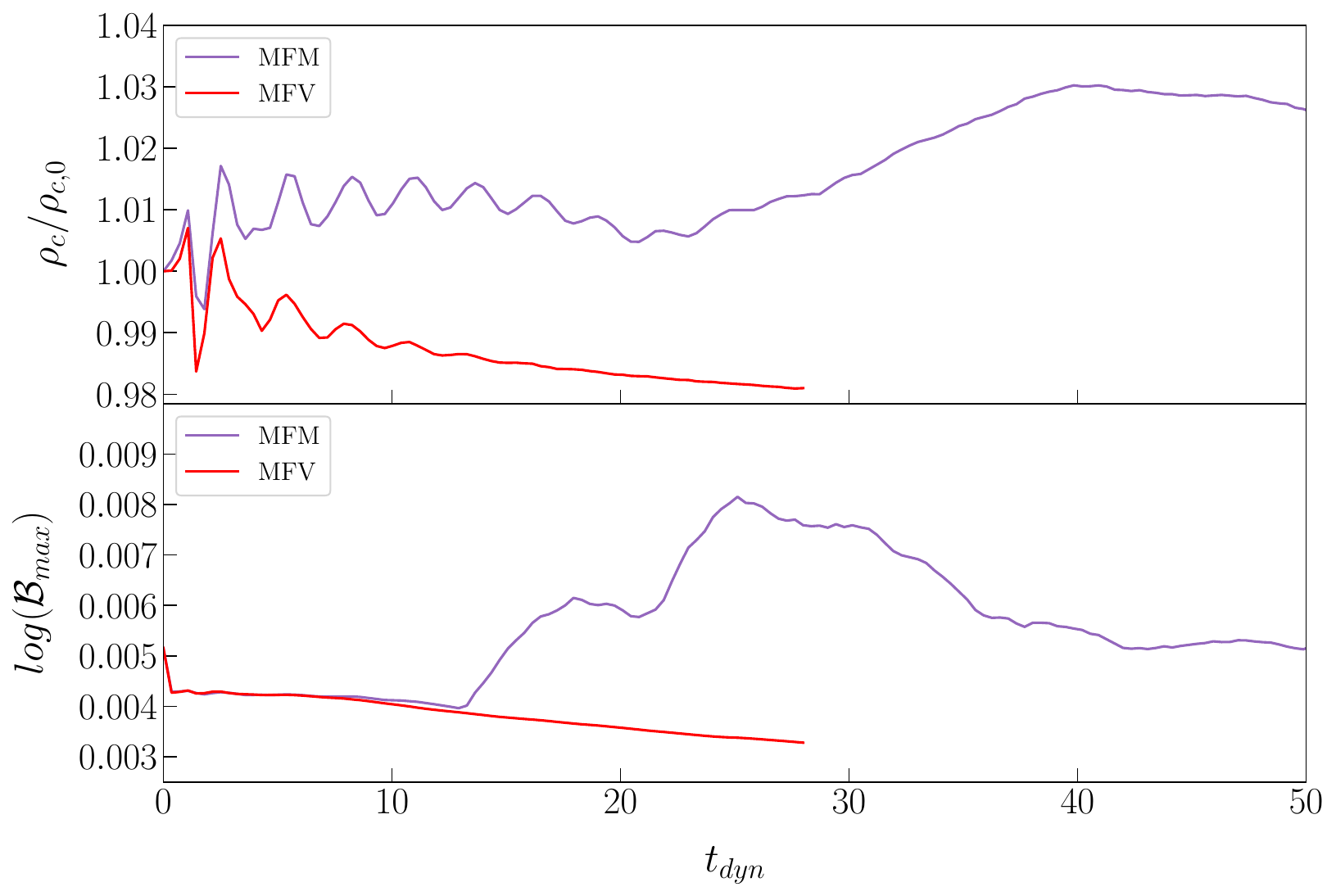}
\caption{Evolution of the central TOV rest mass density (top) and the maximum value of the magnetic field (bottom) for the MFM (purple) and MFV (red) run.}\label{fig:TOVRhoCBMax}
\end{figure}

To better understand the evolution of the system we plot the central rest mass density and maximum magnetic field intensity evolution in Fig. \ref{fig:TOVRhoCBMax}. As mentioned above, in the MFV case, the $\rho_c$ evolution is comparable with the hydrodynamical case, exhibiting a slow decay over time; the magnetic field displays a similar behavior: after an initial drop in the central region, the magnetic field intensity witnesses a slow decay, approaching $\sim30\%$ at the end of the run. As also studied in \citet{Cipolletta2020}, this trend is related to the drop in the central rest mass density and stabilizes with increasing resolution.
The initial part of the MFM evolution can be compared with the corresponding pure hydrodynamics run. However, after $\sim13$ dynamical times, the magnetic field witnesses a topological rearrangement and an increase of its maximum intensity by a factor $\sim1.6$. When using the MFM scheme to evolve an equilibrium configuration, setting the frame velocity in the Riemann problems to cancel out the mass fluxes introduces small perturbations in the fluid velocity. In a pure GRHD scenario, this effect does not pose significant issues, as the quantities readjust correctly without notable consequences. In the GRMHD case, instead, these perturbations significantly affect the evolution of the magnetic fields, leading to the aforementioned increase; as a consequence, the central rest mass density also increase by $\sim2\%$, with a delay. In order to assess the long-term effect of this hitch, we let the system evolve up to $t\approx50t_{dyn}$. Despite the artificial rearrangement of the magnetic field, the configuration does not appear to diverge; the TOV star remains stable and both the $\mathcal{B}$ maximum value and the central density eventually (slowly) decay towards their initial value, albeit with a noisier spatial configuration.

\subsubsection{Magnetized Bondi accretion}\label{sec:BondiAccretion}
As a final test, we consider the magnetized spherical accretion flow onto a non-spinning BH with mass $M_{BH}$, as a GRMHD extension of the test performed in \citet{Lupi2023}. The solution represents the relativistic generalization of the Bondi accretion solution \citep[]{Michel1972,Hawley1984}, with the addition of a radial magnetic field given by
\begin{equation}\label{eq:BondiBField}
    \mathcal{B}^r = \frac{B_0M_{BH}^2}{r^2},
\end{equation}
where $r$ is the coordinate radius and $B_0$ is a free parameter that controls the magnetic field strength. This spatial configuration of $\mathcal{B}$ does not alter the hydrodynamical solution and satisfies the divergence-free constraint (\ref{eq:DivergenceFreeB}).

We recall that the steady-state solution can be written as \citep{LiptaiPrice2019}
\begin{equation}\label{eq:BondiHDSolution}
\begin{split}
    &u^r(r)=\frac{C_1}{r^2\zeta^n(r)};\\
    &\rho(r)=K_0\zeta^n(r);\\
    &\epsilon(r)=n\zeta(r),
\end{split}
\end{equation}
where $\zeta\equiv (\Gamma-1)\epsilon$ and $n\equiv1/(\Gamma-1)$. By assuming a critical radius $r_c$, where
\begin{equation}\label{eq:BondiCriticalRadius}
\begin{split}
    &u_c\equiv u^r(r_c)=\sqrt{\frac{M_{BH}}{2r_c}};\\
    &v_c\equiv v^r(r_c)=\sqrt{\frac{u_c^2}{1-3u_c^2}};\\
    &\zeta_c\equiv \zeta(r_c)=\frac{nv_c^2}{1+n(1-v_c^2)-n^2v_c^2},
\end{split}
\end{equation}
one can determine the coefficients
\begin{equation}\label{eq:BondiC1C2}
\begin{split}
    &C_1=u_cr_c^2\zeta_c^n;\\
    &C_2=(1+(n+1)\zeta_c)^2\left(1-\frac{2M_{BH}}{r_c}+u_c^2\right)
\end{split}    
\end{equation}
and retrieve $\zeta(r)$ by solving
\begin{equation}\label{eq:BondiZetaEquation}
    C_2=(1+(n+1)\zeta(r))^2\left\{1-\frac{2M_{BH}}{r}+[u^r(r)]^2\right\}.
\end{equation}
Using Eq.s (\ref{eq:BondiHDSolution}) and taking into consideration the curved metric tensor, we then compute the Lorentz factor by solving
\begin{equation}\label{eq:BondiLorentzFactor}
    W=\frac{1}{\sqrt{1-\gamma_{ij}v^iv^j}}=\frac{1}{\sqrt{1-\gamma_{ij}(\frac{u^i}{W}+\frac{\beta^i}{\alpha})(\frac{u^j}{W}+\frac{\beta^j}{\alpha})}}
\end{equation}
and the coordinate velocity $\tilde{v}^i=\alpha u^i/W$. Finally we evaluate $D(r)=\sqrt{\gamma}\rho W$.

Unlike \citet{Lupi2023}, here we work with ``Cartesian'' Kerr-Schild (KS) coordinates \citep[]{Kerr1963,Visser2008,Blakely}, as they are horizon-penetrating, i.e. the metric is continuous across the event horizon at $r=2M$. In KS coordinates, the general ($a\neq0$) metric can be expressed as
\begin{equation}\label{eq:KerrSchildMetric}
    g_{\mu\nu}=\eta_{\mu\nu} + \frac{2M_{BH}r^3}{r^4+a^2z^2}l_al_b,
\end{equation}
where $\eta_{\mu\nu}$ is the Minkowski flat metric, $a$ is the spin parameter,
\begin{equation}\label{eq:KerrSchildRadius}
    r(x,y,z)=\sqrt{\frac{R^2-a^2+\sqrt{(R^2+a^2)^2+4a^2z^2}}{2}}
\end{equation}
is the coordinate radius as a function of the spherical radius $R=\sqrt{x^2+y^2+z^2}$ and
\begin{equation}\label{eq:KerrSchildLaLb}
    l_a=\left(1,\frac{rx+ay}{r^2+a^2},\frac{ry-ax}{r^2+a^2},\frac{z}{r}\right).
\end{equation}
In our test problem, we set $a=0$, hence $r=R$. Note that, in this coordinate system, the shift vector $\beta_i=g_{i0}$ differs from zero even for a non-spinning BH, allowing us to directly test the correct implementation of the shift vector-dependent terms in our scheme. 

\begin{figure*}
\begin{center} 
\includegraphics[width=.95\textwidth]{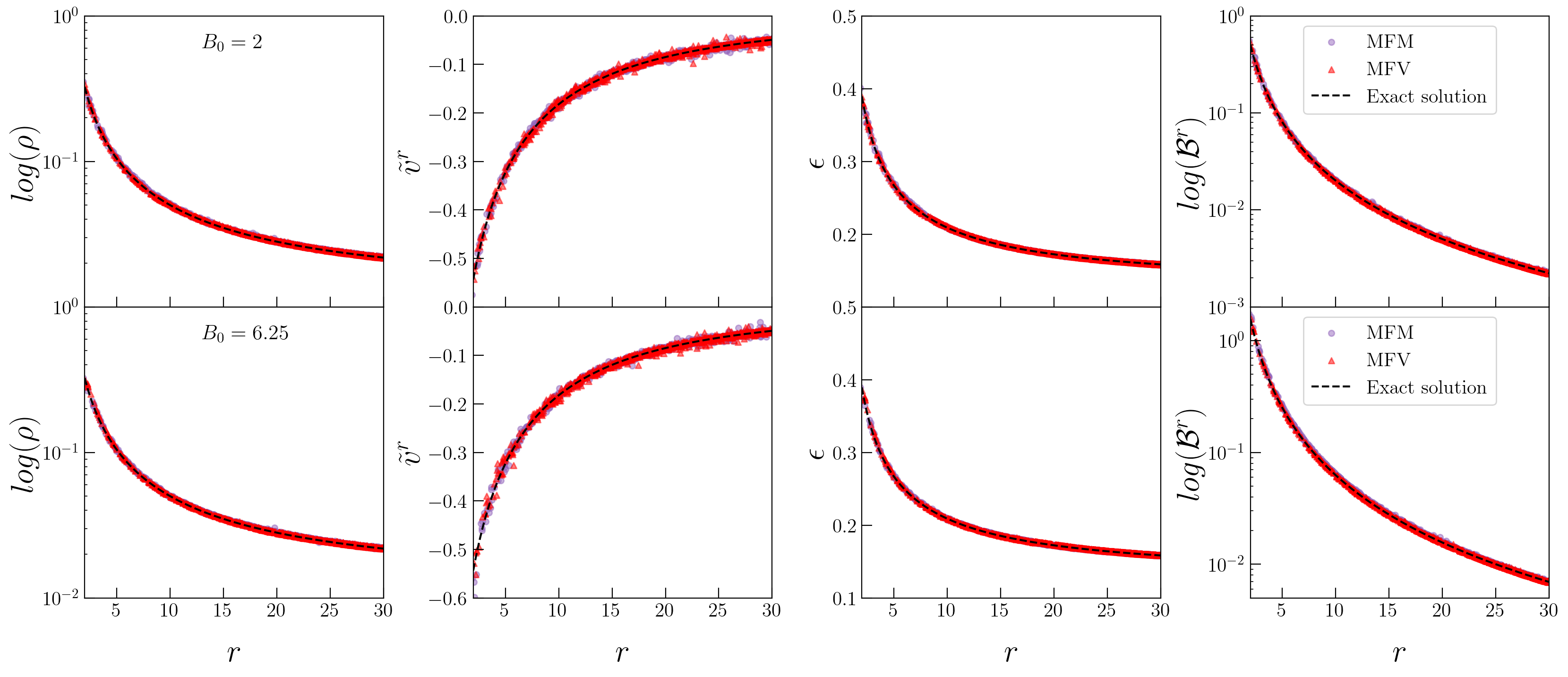}
\end{center}
\caption{Rest-mass density $\rho$, coordinate velocity $\tilde{v}^r$, specific internal energy $\epsilon$ and magnetic field $\mathcal{B}^r$ radial profile of the magnetized Bondi accretion test, performed both in the MFM (purple circles) and MFV (orange triangles) modes, at time $t=100M$. We plot results of the run performed with low (top row) and high magnetization (bottom row). In each panel, we display the exact solution with a dashed line.}\label{fig:Bondi}
\end{figure*}

For our initial condition, we consider a critical radius $r_c=8M$, a $K_0$ value that gives $\rho(r_c)=1/16$ and an adiabatic index $\Gamma=4/3$. We perform runs with two different magnetic field intensities $B_0=(2.0, 6.25)$, yielding a magnetic-to-gas pressure parameter $\beta^{-1}\approx (0.1, 1.1)$ at the critical radius and $\beta^{-1} \approx(4.7,46)$ at the BH horizon. We initially place $\sim2\times 10^7$ particles on a closely packed lattice ranging from $r=1M$ up to $r=100M$ and then stretch their radial position to match the exact density profile. During the evolution, we set the particle excision radius at $r_{exc}=1M$, to prevent them from reaching the central singularity.

In the central regions, where the fluid magnetization is stronger, we notice an overestimation of the gas internal energy when Eq (\ref{eq:GRMHDHyperbolicForm}) is employed, which corrupts the solution \citep{DelZanna2007}. Since we do not expect shocks in this test, i.e. the entropy should remain constant, in these central regions we activate the energy-entropy switch from Eq. (\ref{eq:EntropyAdiabaticEquation}), and recover the internal energy from the entropy function $s\equiv P/\rho^\Gamma$.

To avoid numerical issues inside the event horizon, we also dampen the magnetic field and the velocity components via the function
\begin{equation}\label{eq:SmoothingInsideHorizon}
    \chi =
\begin{cases}
    \exp[-\frac{\sqrt{r_{exc}}(r-r_{EH})^2}{(r_{EH}-0.4)}]; \ \ \ &r \leq r_{EH} \\
    1; &\rm otherwise
\end{cases}
\end{equation}
where $r_{exc}$ and $r_{EH}$ are the excision and the event horizon radii respectively. 

We report in Fig. \ref{fig:Bondi} the results from runs performed in MFV and MFM modes, for both the low and the high magnetization cases. We report radial profiles of the rest-mass density $\rho$, the radial coordinate velocity $\tilde{v}^r$, the specific internal energy $\epsilon$ and the radial component of the conserved magnetic field $\mathcal{B}^r$ at $t=100M$, in the range $r\in[2,30]M$. Our solutions are in very good agreement with the exact solution. We recognize a small scatter in the radial velocity profile (though more moderate with respect to the hydrodynamical case) and a small overestimation of the same quantity close to $r=2M$ in the strong magnetization case.
Thanks to the horizon penetrating metric, our solution is smooth through $r=2M$ and does not show any drop in density or specific internal energy near this radius.
For completeness purposes, in Appendix \ref{AppendixA} we discuss the same problem evolved without the energy-entropy switch.

\section{Discussions and Conclusions}\label{Sec:Conclusion}
In this paper, we have presented a novel GRMHD scheme implemented in the code \texttt{GIZMO}, that works with the MFV and MFM modes, and builds up on the GRHD implementation presented in \citet{Lupi2023}. The implementation incorporates a general relativistic single-state HLL Riemann solver that considers magnetic fields, a more stable reconstruction of the velocity field at the faces between interacting particles and an energy-entropy switch, which is employed when the fluid is supersonic, strongly magnetized and isentropic. In the generalised kick-drift-kick scheme, we introduced an explicit time evolution of the conserved quantities which takes into account the particle volume variation during a timestep; in this way, we consistently pass the local, un-densitized evolved variables to the GRMHD conservative-to-primitive routine. To control the magnetic field divergence errors, we implemented a general relativistic version of the Powell "8-wave" \citep{Powell1999} and of the Dedner hyperbolic divergence cleaning \citep{Dedner2002}. We tested and calibrated these routines and verified that the scheme can evolve the system with reasonably low $\nabla \cdot B$ values.

We performed several benchmark tests, starting with planar relativistic MHD shock tube problems from \citet{Balsara2000} and then testing the magnetic field loop advection and the magnetic rotor setup from \citet{DelZanna2003}, in two dimensions. As more stringent tests, we performed the magnetized cylindrical (2D) and spherical (3D) blast waves, both in the MFV and the MFM code modes, also employing different kernel functions. From this set of multidimensional special relativistic magnetized tests, we conclude that our code can stably evolve various problems yielding good results. We observe no dependence on the geometry of the problem and the magnetic field divergence is always kept under safe levels. However, we recognize some moderate inaccuracies when dealing with violent shocks characterized by a strong density/pressure gradient and high magnetization. These artifacts are enhanced when a sharp particles volume gradient is initialized.

We then challenged our scheme against GRMHD tests, such as the stability of the magnetized TOV and the magnetized spherical accretion onto a non-rotating black hole. In the first one, we confirm that the MFV mode is better at maintaining the magnetic field topology intact, even though the code can preserve the star hydrodynamical equilibrium in both modes. For the second test, we implemented the horizon-penetrating Kerr-Schild coordinate system and, employing the energy-entropy switch, we achieved an accurate steady-state solution.

We finally note that our GRMHD implementation can work with any equation of state and metric provided by the user, while the ideal fluid EoS, the flat and Kerr metrics (both in Boyer-Lindquist and Kerr-Schild coordinates) are already present in the code. Our scheme will be made public in due time.
In forthcoming papers, we will include dynamic metrics \citep[]{Combi2021,Combi2024} and a more sophisticated Riemann solver \citep[]{MignoneBodo2006, MiyoshiKusano2005, Mignone2009}.
\section{Acknowledgments}

Some simulations were run on the Leonardo cluster at CINECA (allocation INF24\_teongrav) thanks to the CINECA-INFN agreement.
The authors thank the reviewer, Daniel J. Price, for his valuable comments and suggestions that improved the quality of the article. 
GF thanks Bruno Giacomazzo for his advice about the TOV equilibrium test, and Andrea Mignone and Vittoria Berta for helpful discussions.
AL acknowledges support from PRIN MUR “2022935STW".
\bibliographystyle{aa}

\bibliography{bibliografia}

\appendix

\section{Magnetized Bondi accretion without energy-entropy switch}\label{AppendixA}
\begin{figure}
\begin{center} 
\includegraphics[width=\columnwidth, trim=0 0 21cm 0,clip]{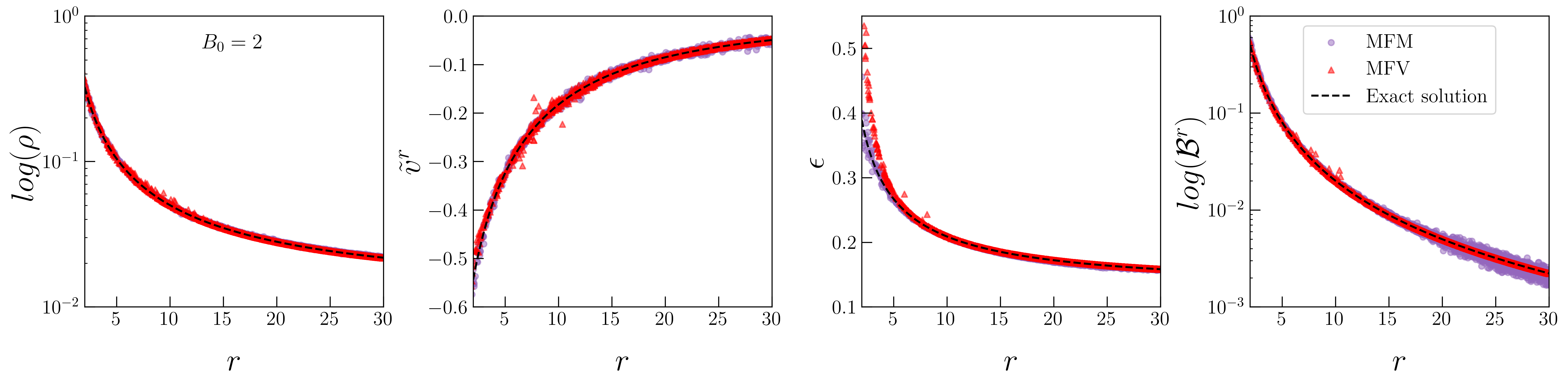}
\includegraphics[width=\columnwidth, trim=21cm 0 0 0,clip]{Plots/Bondi/Bondi_r_NoSwitch_new.png}
\end{center}
\caption{Radial profiles of the rest-mass density $\rho$, the radial coordinate velocity $\tilde{v}^r$, the specific internal energy $\epsilon$, and the radial component of the magnetic field $\mathcal{B}^r$ of the magnetized Bondi accretion test, performed without the energy-entropy switch. We show the results obtained with the MFV (orange triangles) and MFM mode (purple circles), at time $t=10M$ in the low-magnetization case. The exact solution is reported in each panel by a dashed black line.}\label{fig:BondiNoSwitch}
\end{figure}

We report here the results for the magnetized Bondi accretion evolved without the energy-entropy switch discussed in Sec. \ref{sec:C2PConversionWvReconstruction}. The initial setup is identical to the one presented in Sec. \ref{sec:BondiAccretion}. Here, we evolve the conserved total energy $\tau$ through the usual conservation Eq. (\ref{eq:GRMHDHyperbolicForm}), instead of using the entropy conservation (\ref{eq:EntropyAdiabaticEquation}). In this case, we let the system evolve until $t=10M$ and we plot the results in Fig. \ref{fig:BondiNoSwitch}. Despite the earlier time, we already observe a significant increase (1.5x) of the specific internal energy $\epsilon$ at the event horizon, in the MFV run. As mentioned in the main text, this effect is due to an incorrect retrieval of the specific internal energy from the conserved energy, when dealing with supersonic flows, strong gravitational fields and relatively high magnetic field pressures $b^2/2$. Despite this hitch, the rest-mass density, the coordinate velocity, and the magnetic field remain consistent with the expected solution overall, but for a small scatter around $r=10M$. A subtle underestimation of the radial velocity is found near the event horizon, which is a direct consequence of the incorrect internal energy density estimation. The MFM run yields good results overall, with only a larger scatter in the internal energy and the magnetic field when compared with Fig. (\ref{fig:Bondi}).

\section{Monopole test performed with the Powell scheme only}\label{AppendixB}

\begin{figure}
\begin{center} 
\includegraphics[width=\columnwidth]{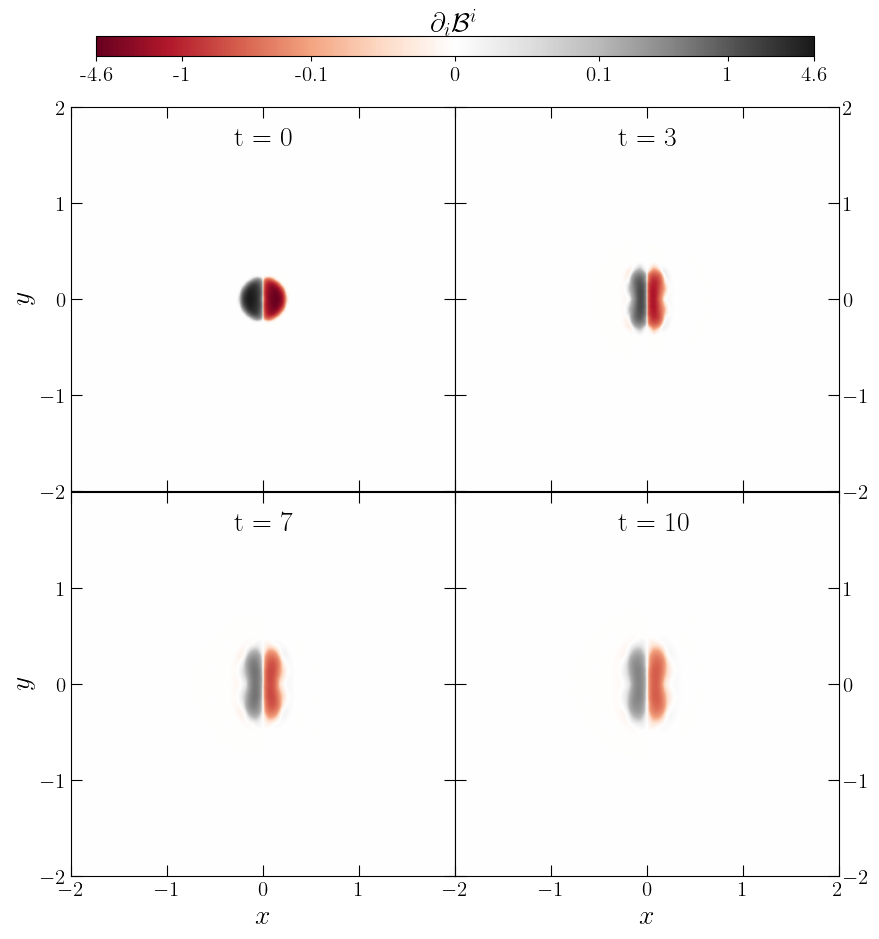}
\end{center}
\caption{Evolution of the monopole test performed with the Powell scheme only. We plot a two-dimensional slice at $z=0$ of the magnetic field divergence at the initial time and at three later times. Note the different color-map scale with respect to Fig. ~\ref{fig:Monopole}.}\label{fig:MonopolePowell}
\end{figure}

To better show how our divergence cleaning scheme handles spurious magnetic field divergences, we performed the monopole test from Sec. \ref{sec:Monopole} employing the Powell terms in Eq. (\ref{eq:PowellSources}) only. The initial setup is analogous to the one presented in the main text.
We let the system evolve until $t=10$ and we plot the resulting $\partial_i \mathcal{B}^i$ in Fig. ~\ref{fig:MonopolePowell}, at four different times. At the beginning of the evolution, the Powell scheme is highly effective and quickly damps part of the monopole. After $t=5$ it starts to slow down due to the less pronounced $\mathcal{B}$ gradients entering Eq.~(\ref{eq:PowellSources}).
By the final time, the monopole is damped by an order of magnitude and, as expected, we see no advection of the divergence over time.
The small $\partial_i\mathcal{B}^i$ diffusion along the y-direction is a consequence of the induction and the moment density equations reacting to the presence of an initial non-zero magnetic field along the x-direction. The overall behavior of this solution was expected; in fact, while the Powell prescription can in principle correct large magnetic field divergences over time, it is better suited to remove the instantaneous formation of spurious errors emerging from the fluxes evaluation. The cleaning of a strong monopole, such as the one in this test, is best handled by the Dedner hyperbolic cleaning.

\end{document}